# Generation of Subfemtosecond Deep and Vacuum UV pulses via Two-Photon Rabi Oscillations in Alkali Atoms or Alkaline Earth Ions

I.R. Khairulin[1,2,*], A.A. Romanov[1], A.A. Silaev[1], M.Yu. Ryabikin[1,2] and V.A. Antonov[1]

[1]A.V. Gaponov-Grekhov Institute of Applied Physics of the Russian Academy of Sciences, Nizhny Novgorod, Russia

[2]Lobachevsky State University of Nizhny Novgorod, Nizhny Novgorod, Russia

A method is proposed for the formation of femto- and subfemtosecond pulses of the deep ultraviolet and vacuum ultraviolet radiation via generating the third harmonic of femtosecond laser pulses during their resonant interaction with alkali atoms or alkaline earth ions. The pulse formation occurs due to two-photon Rabi oscillations between quasi-equidistant energy levels of atoms or ions. The duration of the generated third harmonic pulse is several times shorter than far from resonance, while the generation efficiency is up to 3-4 orders of magnitude higher.

The end of the 20th – beginning of the 21st century were marked by the rapid development of methods for generating progressively shorter pulses of electromagnetic radiation. By the turn of the new millennium, laser sources provided the generation of near-infrared (IR) pulses with a duration of 4–5 femtoseconds, shorter than two optical cycles [1, 2], and stabilization of the optical carrier-envelope phase [3, 4], see also [5, 6]. In the 2000s, high-harmonic generation (HHG) of the laser field made it possible to obtain attosecond pulses in the extreme ultraviolet (XUV) and soft X-ray ranges [7–10] (for a review of the current state of research in this area, see [11–13]). Together, laser sources and equipment implementing HHG formed the toolkit of attosecond physics, a science focused on studying and controlling the dynamics of electronic processes in atoms, molecules, and solids on their intrinsic time scales [6, 13–16].

In recent years, another class of sources of ultrashort electromagnetic pulses has been rapidly developing, which provide deep ultraviolet (DUV) and vacuum ultraviolet (VUV) pulses with photon energy of ~ 5–20 eV and duration of ~ 1 fs. This range of photon energy covers excitation energies and ionization potentials of different neutral media, which opens up possibilities for their most efficient ionization and control of localized electron wave packets. In attosecond experiments, these pulses can act as a pump with a shorter duration than IR laser pulses [17]. In addition, they can be used in attosecond transient absorption spectroscopy [14] of media with a low ionization potential, where the use of pulses produced via HHG for these purposes is problematic. The most widely used method for generating short pulses in the DUV and VUV ranges is four-wave mixing accompanied by phase self- and cross-modulation of the spectral components of laser radiation and the generation of a supercontinuum in gas cells [17-21] or in gas-filled hollow-core fibers [22-26]. The duration of the generated pulses in this case can reach 2–3 fs (less than 1 fs in the limiting case), and their energy typically varies from fractions to tens of µJ. Additionally, few-cycle pulses of DUV and VUV radiation can be produced by multiwave mixing during gas ionization by a two-component laser field [27].

In this Letter, we propose a new method for generating femto- and subfemtosecond pulses in DUV and VUV ranges based on the third harmonic generation of femtosecond laser pulses in a resonant medium consisting of alkali atoms or alkaline earth ions, which have excited energy levels structure similar to alkali atoms. Such media are characterized by a ladder structure of levels, including the ground, lower excited, and two higher excited states (see Fig. 1). It is shown that there is an optimal area of the laser pulse at which two-photon Rabi oscillations [28, 29] between the ground and excited states of alkali atoms or alkaline earth ions lead to the generation of an ultrashort (down to subfemtosecond duration) third harmonic pulse in the DUV or VUV range. The duration of the generated pulse is several times shorter than far from the resonance (and can be as short as a couple of carrier frequency oscillations), while the generation efficiency is significantly (up to 3-4 orders of magnitude) higher.

Let us illustrate the method using the example of the Na atom. The diagram of its energy levels is shown in Fig. 1(a). The atom is irradiated by an intense linearly polarized laser field, which has the form

$$E_L(t) = \tilde{E}_L(t)\cos(\Omega t), \tag{1}$$

where $\tilde{E}_L(t)$ is the pulse envelope given by $\tilde{E}_L(t) = E_0 \exp\left(-2\ln(2)t^2/\Delta t_p^2\right)$; $E_0$ and $\Omega$ are the amplitude and carrier frequency of the laser field, and $\Delta t_p$ is the pulse duration at half-maximum of the intensity. The field frequency is assumed to be equal to the frequency $\Omega = \omega_{21}$ of the transition from the ground state of the atom, $|1\rangle$, to the first excited state, $|2\rangle$ (lower red arrow in Fig. 1(a)). For the Na atom, $|1\rangle = |2p^63s\rangle$, $|2\rangle = |2p^63p\rangle$, and $\hbar\omega_{21} \simeq 2.10\,\text{eV}$ ($\hbar$ is the reduced Planck's constant) [30]. Due to the proximity of the transition frequencies $|1\rangle \leftrightarrow |2\rangle$, $|2\rangle \leftrightarrow |3\rangle$, and $|2\rangle \leftrightarrow |4\rangle$, where $|3\rangle = |2p^65s\rangle$ and $|4\rangle = |2p^64d\rangle$, the same field also resonantly populates the states $|3\rangle$, and $|4\rangle$. Since a linearly polarized field excites an atom without changing the projection $M$ of the atomic angular momentum onto the direction of field polarization, in all the excited states of the atom under consideration, $M = 0$, as in the ground state. Note that for a given laser field frequency, the third harmonic radiation has a photon energy $\hbar\omega_{3H} \simeq 6.31\,\text{eV}$ and a wavelength in vacuum of 196 nm, which corresponds to the DUV range.

The intensity $I_3$ of the generated third harmonic of the fundamental frequency in an optically thin medium of sodium atoms is determined by the squared modulus of the complex amplitude of the dipole acceleration at the third harmonic frequency, $I_3 \propto \left|\tilde{\tilde{d}}_{3H}\right|^2$. The consideration of the effects of propagation is beyond the scope of this article, thus, in what follows, when omitting the dimensional factor, we will assume $I_3 = \left|\tilde{\tilde{d}}_{3H}\right|^2$. To calculate the dipole acceleration of the Na atom in the field (1), two approaches are used in this paper: (a) a direct numerical integration of the three-dimensional time-dependent Schrödinger equation (TDSE) in the single active electron approximation using the split-step method in spherical coordinates [31] and the effective ion potential calculated within the framework of the density functional theory [32] (see Section 1 in Supplemental Material), and (b) an analytical solution of the TDSE in a four-level model that takes into account only resonant states $|1\rangle - |4\rangle$ and uses the state parameters corresponding to the approach (a) (see Section 2 and Table 1 in Supplemental Material).

Let us start with an analytical solution. For simplicity, we neglect the insignificant differences in the frequencies of transitions $|1\rangle \leftrightarrow |2\rangle$ and $|2\rangle \leftrightarrow |3\rangle, |4\rangle$ (see Fig. 1 and Tables 1, 2 in Supplemental Material) and assume $\Omega = \omega_{21} = \omega_{32} = \omega_{42}$, where $\omega_{ij}$ is the $|i\rangle \leftrightarrow |j\rangle$ transition frequency. In this case, the time dependence of the third harmonic intensity is given by (in atomic units) (see Appendix A)

$$I_3(t) = \alpha \left(\tilde{E}_L/E_0\right)^4 R(D\xi), \tag{2}$$

where

$$\alpha = \left[\frac{9}{32} \cdot \frac{d_{23}^2 + d_{24}^2 - 2d_{12}^2}{D} \cdot (d_{12}E_0)^2\right]^2,$$

$$R(D\xi) = \sin^2(D\xi)\left[1 + \frac{2d_{12}^4 - (d_{23}^2 + d_{24}^2)^2}{2D^2(d_{23}^2 + d_{24}^2 - 2d_{12}^2)}\sin^2(D\xi/2)\right]^2, \quad (3)$$

$d_{ij}$ ($ij = 12, 23, 24$) is the dipole moment of the transition $|i\rangle \leftrightarrow |j\rangle$, $D = \sqrt{d_{12}^2 + d_{23}^2 + d_{24}^2}/2$, $\xi = \int_{-\infty}^{t} \tilde{E}_L(t')dt'$, and $D\xi$ is the local value of the area under the laser pulse envelope. Note that the solution (2), (3) was obtained under the assumption that the Rabi frequencies at the transitions under consideration, $d_{ij}E_0$, $\{i,j\} = \{1,2; 2,3; 2,4\}$, are significantly lower than the laser field frequency $\Omega$: $d_{ij}E_0/\Omega \ll 1$. According to (2), the time dependence of the third harmonic intensity is determined by the product of the fourth power of the laser pulse electric field envelope $\tilde{E}_L$ and the function $R(D\xi)$. The dependence on $\xi$ of the function $R(D\xi)$ is a periodic sequence of beats, which originate from two-photon Rabi oscillations between the resonant atomic states and interference between the different paths of the third harmonic emission (see Appendix A). In turn, the envelope of the laser pulse is nonzero within the interval $0 < \xi < S_p$, where $S_p = \xi(t \to \infty)$ (thus, the total area of the laser pulse is equal to $DS_p$). It selects in $R(D\xi)$ only the beats that fit into this interval, see Fig. 2(a). For the laser pulse with a Gaussian envelope shape (1), $S_p$ is determined by

$$S_p = \sqrt{\frac{\pi}{2\ln 2}} E_0 \Delta t_p. \quad (4)$$

For the optimal value of the total laser pulse area, $DS_p = 1.59\pi$, the laser envelope covers three bursts in $R(\xi/S_p)$, see Fig. 2(a). The first and third bursts are located on the leading and trailing edges of the laser pulse, respectively, where the value of $(\tilde{E}_L/E_0)^4$ is small, while the second burst is in the vicinity of the maximum of $(\tilde{E}_L/E_0)^4$. As a result, an isolated pulse is produced in the time dependence $I_3(t)$ of the third harmonic intensity (see Fig. 2(b)), along with two side bursts of equal intensity approximately 2.8 times lower than the intensity of the main burst. The duration of the produced third harmonic pulse $\Delta t_{3H}$ is 2.7 times shorter than the similar duration in the case of a medium with nonresonant cubic nonlinearity. In the latter case, a laser pulse of Gaussian shape (1) with a duration $\Delta t_p$ generates third harmonic radiation of duration $\Delta t_{3H}^{(nonres)} = \Delta t_p/\sqrt{3}$. In the case shown in Fig. 2(b), which corresponds to $\Delta t_p = 20$ fs and the laser intensity $I_0 = 2.9 \times 10^{11}$ W/cm$^2$, we get $\Delta t_{3H} \approx 4.3$ fs, while $\Delta t_{3H}^{(nonres)} \approx 11.5$ fs. The optimal area of the laser pulse, $DS_p = 1.59\pi$, corresponds to approximately one and a half Rabi oscillations between the states $|1\rangle$ and $|2\rangle$ and significant population of the states $|3\rangle$ and $|4\rangle$ (see Supplemental Material, Fig. S4(a)). The time dependences of the third harmonic signal in the cases of weak atomic excitation ($DS_p \ll \pi$) and multiple Rabi oscillations between the resonant states ($DS_p \gg \pi$) are discussed in Appendix B, as well as in Supplemental Material, Section 3.

Let us introduce the compression coefficient $\beta$ as the ratio of the duration of the laser field pulse $\Delta t_p$ to the duration of the generated third harmonic pulse $\Delta t_{3H}$:

$$\beta = \Delta t_p / \Delta t_{3H}. \quad (5)$$

The value $\beta$ shows how many times the generated third harmonic pulse is shorter than the generating pulse of the laser field. According to the analytical solution (2), (3), the shape of the third harmonic pulse is determined exclusively by the area of the laser pulse $DS_p$. As a result, with an optimal choice of the laser intensity, the compression coefficient does not depend on its duration and for Na atoms equals $\beta \simeq 4.6$.

The optimal laser pulse area, as well as the value of the compression coefficient discussed above, were obtained on the basis of the analytical solution (2), (3), which does not take into account (a) the differences between the frequencies of the transitions $\omega_{21}$, $\omega_{32}$, and $\omega_{42}$, (b) atomic ionization, (c) transitions to other excited states, and (d) is valid for not-too-intense laser field ($d_{ij}E_0/\Omega \ll 1, \{i,j\} = \{1,2; 2,3; 2,4\}$). In order to take into account the above factors, we solved the TDSE for the Na atom and the Mg$^+$ ion from the first principles (see Supplemental Material, Section 1). The Mg$^+$ ion has an energy structure similar to that of the Na atom, with the resonance states being $|2p^63s\rangle = |1\rangle$, $|2p^63p\rangle = |2\rangle$, $|2p^63d\rangle = |3\rangle$, and $|2p^64s\rangle = |4\rangle$, but with higher transition frequencies (see Fig. 1(b)). The interest in using Mg$^+$ ions is due to the high photon energy, 13.3 eV, of the generated third harmonic field, which cannot be produced by nonlinear crystals [33], and the possibility to use of the third harmonic of a titanium-sapphire laser radiation as the resonant driving field (see Table 2 in Supplemental Material).

The results of the numerical *ab initio* TDSE solution are presented in Fig. 3. Figure 3(a) shows the values of the peak intensity $I_0$ of the laser pulse for its different durations in the range 5 fs $\leq \Delta t_p \leq$ 60 fs, which minimize the duration of the third harmonic pulse under the condition of small intensities of side bursts (see Supplemental Material, Sections 3, 4). Figure 3(b) shows the values of the compression coefficient corresponding to these combinations of $\Delta t_p$ and $I_0$. The blue and red colors correspond to the Na atom and the Mg$^+$ ion, respectively. In the latter case, according to the analytical solution, the optimal value of the laser pulse area is $DS_p = 3.27\pi$ (see Supplemental Material, Section 4), and the third harmonic compression coefficient is $\beta \simeq 5.4$. As follows from Fig. 3, the generation of a shortened pulse of third harmonic predicted by the analytical solution (2), (3) is also observed in the *ab initio* solution of TDSE. The results of both approaches are qualitatively similar, and the differences between them are easily explained (see Appendix C). Here, we only note that for a short duration and high peak intensity of the laser pulse, the difference between the above approaches is caused by the ionization of the atom (ion), the convergence of the Rabi frequencies with the laser field frequency, and the involvement of an increasing number of bound states in the interaction with the field. For the long duration and low intensity of the laser pulse, the difference is mainly due to the non-equidistant energies of the resonant states under consideration.

As can be seen from Fig. 3(b), the proposed mechanism of third harmonic generation allows obtaining pulses of DUV and VUV radiation down to subfemtosecond duration. In the case of the Na atom, the use of a resonant laser pulse with an intensity of $I_0 = 2.5 \times 10^{13}$ W/cm$^2$ and duration of $\Delta t_p = 5$ fs allows obtaining a pulse of DUV radiation with a photon energy of 6.3 eV and a wavelength of 196 nm with a duration of $\Delta t_{3H} = 1.4$ fs (see Fig. 4(a)), which is about two cycles of carrier-frequency field oscillations. In the case of the Mg$^+$ ion, a resonant laser pulse with the same duration and intensity allows one to obtain a VUV radiation pulse with a photon energy of 13.3 eV and a wavelength of 93.2 nm with a duration of $\Delta t_{3H} = 0.97$ fs (Fig. 4(b)) (about three cycles of field oscillations at the carrier frequency).

The regime of formation of ultrashort pulses of third harmonic radiation investigated above is robust to changes in the laser field intensity by 2–3 times (see Supplemental Material, Section 5). At the same time, the permissible frequency detuning of the laser field from the resonance with the atomic transition is of the order of the Rabi frequency of the laser pulse and increases with decreasing its duration. For instance, for the Mg$^+$ ion and a laser pulse duration of 5 fs, the permissible range of carrier frequency tuning is ±10% in the vicinity of 4.77 eV (in a real ion, the transition energy is 4.43 eV).

Finally, we note that the efficiency of generating third harmonic radiation in resonance with a cascade two-photon transition is significantly greater than far from it; the difference can be up to 3–4 orders of magnitude (see Supplemental Material, Section 5).

To conclude, we proposed a method for the formation of femto- and subfemtosecond pulses in DUV and VUV ranges in the process of generating the third harmonic of a femtosecond laser pulse during its resonant interaction with alkali atoms or alkaline earth ions. It is shown that for an arbitrary duration of a laser pulse with an optimal choice of its area, as a result of excitation of two-photon Rabi oscillations between quasi-equidistant lower energy levels of an atom or ion, the generated pulse of third harmonic radiation is 2–4 times shorter than in a non-resonant medium with cubic nonlinearity. In its turn, the efficiency of third harmonic generation in this case is significantly higher than far from resonance. This third harmonic generation regime is robust to a 2–3-fold change in the laser field intensity, as well as to its detuning from the resonance by an amount of the Rabi frequency of the laser pulse. The proposed method doesn't rely on macroscopic effects and requires just moderate laser intensity of the order of $10^{11}$-$10^{13}$ W/cm$^2$. When using laser pulses with a duration of ~ 5 fs, the third harmonic pulses can have a subfemtosecond duration. In particular, in a medium of Mg$^+$ ions, using the third harmonic of Ti:Sa laser radiation as a driving field, a pulse of 0.97 fs duration at a wavelength of 93.2 nm is generated. In a medium of Na atoms, the possibility of generating a pulse of 1.4 fs duration at a wavelength of 196 nm is demonstrated. Such pulses can be used for femto- and subfemtosecond metrology and coherent control, including chemical reaction control.

We acknowledge support by the Russian Science Foundation (grant No. 22-12-00389).

**Appendix A. Dipole moment of the four-level system at the frequency of the third harmonic**

In the four-level model, the wavefunction of the outer atomic electron has a form

$$|\Psi\rangle = \sum_{k=1}^{4} a_k(t)|k\rangle. \tag{A1}$$

In the following, we assume that the eigenstates $|k\rangle$ do not depend on time, and the whole time dependence of the wavefunction $|\Psi\rangle$ is due to their probability amplitudes $a_k(t)$.

For a not-too-strong multicycle laser field (1), the solution for $a_k(t)$ has a form of Fourier series in laser field frequency $\Omega$ with slowly varying expansion coefficients:

$$a_k(t) = \sum_{n=-\infty}^{\infty} a_{k,n}(t)e^{-in\Omega t}, \ k=1,2,3,4, \quad \left|\frac{d}{dt}a_{k,n}(t)\right| \ll \Omega|a_{k,n}(t)|. \tag{A2}$$

Then, the slowly varying amplitude of the induced atomic dipole moment at the frequency $3\Omega$ of the third harmonic of the laser field (1) has a form

$$\tilde{d}_{3H} = \sum_{n=-\infty}^{\infty} \sum_{k,s=1}^{4} d_{sk} a_{k,n}^* a_{s,n+3}. \tag{A3}$$

The terms in (A3) correspond to the transitions between different Fourier components of the atomic states and represent different paths of the third harmonic generation.

For a relatively weak, $d_{ij}E_0/\Omega \ll 1$, resonant laser field, $\Omega = \omega_{21} = \omega_{32} = \omega_{42}$, with a slowly varying amplitude, $|d\tilde{E}_L/dt| \ll \Omega \tilde{E}_L$, the amplitudes of the atomic states $|1\rangle$–$|4\rangle$, which satisfy the initial conditoin $a_1(t=-\infty)=1$, $a_k(t=-\infty)=0$ for $k=2, 3$, and 4, are dominated by (see Supplemental Material, Section 2)

$$a_1(t) \simeq a_{1,0}(t) = 1 - \frac{d_{12}^2}{2D^2}\sin^2[D\xi(t)/2],$$

$$a_2(t) \simeq a_{2,1}(t)\exp(-i\Omega t) = -i\frac{d_{12}}{2D}\sin[D\xi(t)]\exp(-i\Omega t),$$

$$a_3(t) \simeq a_{3,2}(t)\exp(-i2\Omega t) = -\frac{d_{12}d_{23}}{2D^2}\sin^2[D\xi(t)/2]\exp(-i2\Omega t),$$ (A4)

$$a_4(t) \simeq a_{4,2}(t)\exp(-i2\Omega t) = -\frac{d_{12}d_{24}}{2D^2}\sin^2[D\xi(t)/2]\exp(-i2\Omega t).$$

The other terms in Fourier series (A2) are much smaller compared to those entering Eqs. (A4) and can be expressed through the latter via the perturbation theory. In this case, the major contribution to $\tilde{d}_{3H}$ is given by the terms

$$\tilde{d}_{3H} \simeq d_{12}a_{1,-2}^*a_{2,1} + d_{12}a_{1,0}^*a_{2,3} + \left[d_{23}a_{3,2} + d_{24}a_{4,2}\right]a_{2,-1}^* \qquad (A5)$$

(see Supplemental Material, Section 2), while the remaining terms in (A3) either compensate each other or are small quantities compared to (A5). Within the perturbation theory, the terms in (A5) are expressed through the dominating Fourier components of the amplitudes of the eigenstates (see Eqs. (A4)) as follows:

$$d_{12}a_{1,-2}^*a_{2,1} = \frac{d_{12}^2\tilde{E}_L^2}{16\Omega^2}d_{12}a_{1,0}^*a_{2,1}$$

$$d_{12}a_{1,0}^*a_{2,3} = \frac{d_{12}}{4\Omega}\left(\tilde{E}_L + \frac{1}{i2\Omega}\frac{d\tilde{E}_L}{dt}\right)(d_{23}a_{3,2} + d_{24}a_{4,2})a_{1,0}^* - \frac{(d_{23}^2 + d_{24}^2 - d_{12}^2)\tilde{E}_L^2}{16\Omega^2}d_{12}a_{2,1}a_{1,0}^*$$

$$\left[d_{23}a_{3,2} + d_{24}a_{4,2}\right]a_{2,-1}^* = \frac{(d_{23}^2 + d_{24}^2 - d_{12}^2)\tilde{E}_L^2}{16\Omega^2}(d_{23}a_{3,2} + d_{24}a_{4,2})a_{2,1}^* -$$

$$- \frac{d_{12}}{4\Omega}\left(\tilde{E}_L + \frac{1}{i2\Omega}\frac{d\tilde{E}_L}{dt}\right)(d_{23}a_{3,2} + d_{24}a_{4,2})a_{1,0}^*$$ (A6)

As seen from (A6), the terms linear in the field and its time derivative exactly compensate each other in (A5), while summing the remaining terms results in

$$\tilde{d}_{3H} \simeq -\frac{(d_{23}^2 + d_{24}^2 - 2d_{12}^2)\tilde{E}_L^2}{16\Omega^2}\left\{d_{12}a_{2,1}a_{1,0}^* - \frac{d_{23}^2 + d_{24}^2 - d_{12}^2}{d_{23}^2 + d_{24}^2 - 2d_{12}^2}(d_{23}a_{3,2}a_{2,1}^* + d_{24}a_{4,2}a_{2,1}^*)\right\}. \qquad (A7)$$

Multiplying (A7) with $-9\Omega^2$, which originates from the second time derivative of $\exp(-i3\Omega t)$, and taking its square modulus leads to (2) and (3) with $R(D\xi)$ corresponding to the function in curly brackets in Eq. (A7).

Accordingly, $R(D\xi) = 0$ either (a) if all the terms (A5) are equal to zero, which happens when the intermediat state $|2\rangle$ is depopulated, $a_2(t) \propto a_{2,1}(t) = 0$, see Eqs. (A7), (A4), or (b) if the terms in (A5) compensate each other, which happens when the function in square brackets in (3) is equal to zero.

**Appendix B. Time dependence of the third harmonic signal for different values of the laser pulse area**

As follows from (2) and (3), if the laser pulse is too short and/or low-intensity, so that its total area $DS_p \ll \pi$, then the excitation of the atom turns out to be insignificant, the intensity of the third harmonic pulse is small, and its time envelope is $I_3 \propto \tilde{E}_L^4(t)\xi^2(t)$.

On the other hand, if the laser pulse is too long and/or highly intense, and its total area $DS_p \gg \pi$, then according to (2), (3), the time dependence of the third harmonic intensity will be a sequence of short beats enclosed under the envelope $(\tilde{E}_L/E_0)^4$. In this case, multiple Rabi oscillations will occur between the resonance states $|1\rangle - |4\rangle$ of the atom. However, a comparison with the numerical solution of the TDSE from first principles shows that the described case is not feasible in practice, since a high-intensity laser pulse rapidly ionizes the atom. Consequently, most of the bursts on the trailing edge of the pulse sequence predicted by solution (2), (3) are suppressed, and the pulse sequence is shortened. In the limiting case, this can lead to the formation of a single third harmonic pulse, which is extremely short compared to the duration of the laser pulse (1). However, due to the rapid, almost complete ionization of the atom, in this regime, the resonant interaction of the laser field with sodium atoms is suppressed, and the efficiency of generating a third harmonic pulse, characterized by the ratio of the harmonic intensity to the laser field intensity (1), turns out to be low.

At the same time, there is an optimal value of the total area of the laser pulse $DS_p \sim \pi$, at which the generation of a single short pulse of third harmonic occurs without overwhelming ionization of the atom and, as a consequence, with a significantly higher efficiency than in the case of $DS_p \gg \pi$. In particular, for Na atoms, the optimal value of $DS_p$ is $1.59\pi$.

**Appendix C. Discussion of the differences between analytical and numerical solutions of the TDSE**

Let us consider the results for the Na atom (blue dashed lines and blue asterisks in Fig. 3). As follows from Fig. 3(a), for a sufficiently long laser pulse, the optimal value of its intensity, obtained on the basis of numerical calculations, is close to but slightly exceeds the value predicted by the analytical solution (2), (3). Moreover, as the laser pulse shortens, the optimal value of its intensity increases faster than predicted by the analytical solution. This is the result of a slight difference between the frequencies of transitions $|1\rangle \leftrightarrow |2\rangle$ and $|2\rangle \leftrightarrow |3\rangle, |2\rangle \leftrightarrow |4\rangle$ which is present in both the real and the model Na atom (see Table 1 in Supplemental Material). In particular, for the model sodium atom $(\omega_{32} - \omega_{21})/\omega_{21} \simeq -0.0251$ and $(\omega_{42} - \omega_{21})/\omega_{21} \simeq 0.0603$. As a result, to achieve the optimal degree of atomic excitation, it is necessary to use a more intense laser pulse than is predicted by the analytical solution. Moreover, the greater the intensity of the laser field, the greater the degree of atomic ionization, which leads to the need to use an even more intense laser field to achieve the optimal excitation degree. In this case, a shortened pulse of the third harmonic is formed not only due to the induced coherence between the resonant bound states but also due to the appearance of coherence at other transitions, as well as the shortening of the third harmonic envelope due to rapid ionization. A small difference in the frequencies of the resonant transitions also leads to a slightly smaller, but almost constant, value of the compression coefficient $\beta$ in the large $\Delta t_p$ region (see blue asterisks in Fig. 3(b)), whereas with a decrease in $\Delta t_p$, the dependence of the compression coefficient $\beta$ on $\Delta t_p$ becomes nonmonotonic due to a change in the shape of the third harmonic pulse caused by atomic ionization, as well as the excitation of nonresonant atomic states.

Next, let us consider the results for the Mg$^+$ ion (red lines and red circles in Fig. 3). In the Mg$^+$ ion, the resonance states are $|2p^63s\rangle = |1\rangle$, $|2p^63p\rangle = |2\rangle$, $|2p^63d\rangle = |3\rangle$, and $|2p^64s\rangle = |4\rangle$, see Fig. 1(b). For the model Mg$^+$ ion, the resonant laser frequency is $\hbar\Omega = \hbar\omega_{21} \simeq 4.77$ eV, while $(\omega_{32} - \omega_{21})/\omega_{21} \simeq -0.0063$ and $(\omega_{42} - \omega_{21})/\omega_{21} \simeq -0.1006$ (see Table 2 in Supplemental Material).

Note that in this case, the resonant frequency of the laser field, $\Omega = \omega_{21}$, is less than ⅓ of the ionization potential of $Mg^+$ ($I_p$ = 15.63 eV, see Table 2 in Supplemental Material). As a result, in addition to the mechanism of third harmonic generation through two-photon cascade excitation for $Mg^+$ ions described above, there is another generation path, namely, – through cascade three-photon excitation of the highly excited states $|2p^6 6p\rangle$, $|2p^6 6f\rangle$, $|2p^6 7p\rangle$, and $|2p^6 7f\rangle$. However, as can be seen from Fig. 3(a), an estimate of the optimal ratio between the intensity and duration of the laser pulse, $DS_p = 3.27\pi$, obtained on the basis of the analytical solution (2), (3), in this case is in good agreement with the dependence $I_0(\Delta t_p)$, found based on the *ab initio* solution of the TDSE, over the entire considered range of laser pulse durations. This may indicate that the third harmonic generation as a result of two-photon cascade excitation of the $Mg^+$ ion dominates over its generation through three-photon excitation. This can be explained by (a) strong single-photon ionization of the highly excited states, (b) the smallness of the dipole moments of transitions from the states $|1\rangle$, $|3\rangle$, and $|4\rangle$ to the highly excited states compared to the dipole moments of transitions $|1\rangle \leftrightarrow |2\rangle$, $|2\rangle \leftrightarrow |3\rangle$, and $|2\rangle \leftrightarrow |4\rangle$ (see Table 3 in Supplemental Material), as well as (c) strong dynamic Stark shift in the energies of the highly excited states. Finally, as in the case of the Na atom, because of small differences in the frequencies of the resonant transitions, the compression coefficient $\beta$ in the medium of $Mg^+$ ions, found based on the numerical solution of the TDSE from first principles, on average turns out to be somewhat smaller than predicted by the analytical solution (2), (3).

*Corresponding author: khairulinir@ipfran.ru

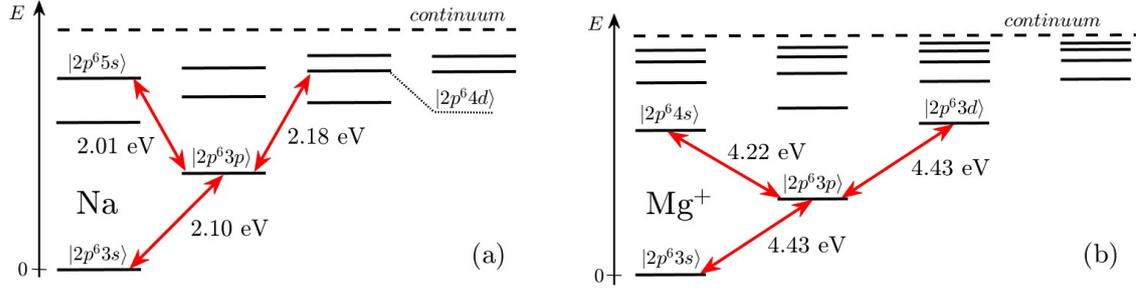

FIG. 1. (Color online) Energy level diagram of (a) Na atom and (b) Mg$^+$ ion. Red arrows indicate transitions involved in two-photon Rabi oscillations; transition energies taken from [30] are indicated next to the arrows.

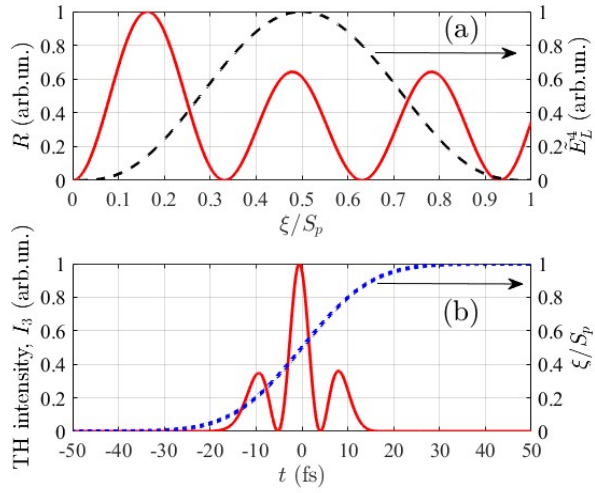

FIG. 2. (Color online) (a) Normalized dependences of the function $R$, see (3) (left axis, red solid line), and the fourth power of the slowly varying amplitude of the Gaussian laser pulse (right axis, black dashed line) on the instantaneous value of the normalized area under the laser pulse. (b) Time dependences of the normalized intensity of the third harmonic (left axis, red solid line) and the normalized instantaneous area under the laser pulse (right axis, blue dotted line). The figures are drawn based on the analytical solution (2), (3) for the Na atom, $\hbar\Omega = 1.99$ eV, $DS_p = 1.59\pi$. Figure (b) corresponds to $\Delta t_p = 20$ fs.

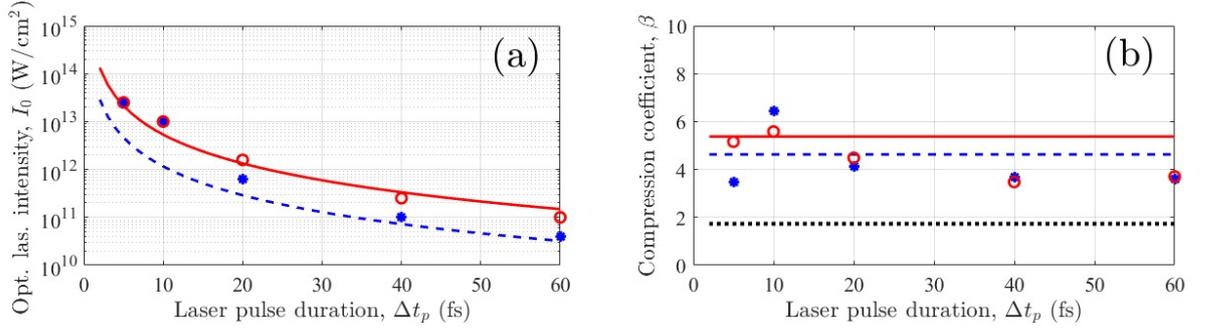

FIG. 3. (Color online) Dependences of (a) the optimal value of the laser pulse intensity $I_0$ and (b) the corresponding compression coefficient $\beta$ on the laser pulse duration $\Delta t_p$. The lines correspond to the analytical solution (2), (3); the markers correspond to the numerical solution of the TDSE. The blue dashed line and asterisks correspond to the Na atom ($\hbar\Omega = 1.99$ eV, $DS_p = 1.59\pi$); the red solid line and circles correspond to the Mg$^+$ ion ($\hbar\Omega = 4.77$ eV, $DS_p = 3.27\pi$); the black dotted line shows the compression coefficient for a nonresonant medium with cubic nonlinearity, $\beta_{nonres} = \sqrt{3}$.

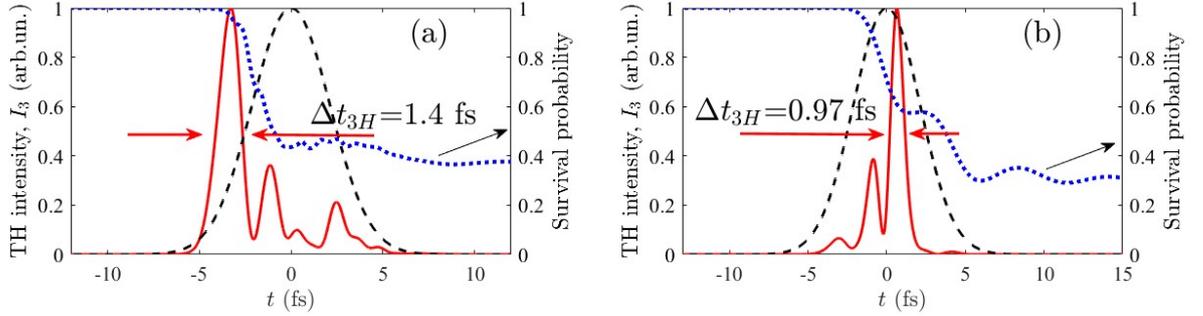

FIG. 4. (Color online) Time dependences of the third harmonic intensity (left axis, red solid line) and ionization probability (right axis, blue dashed line) obtained based on the numerical solution of TDSE. Figure (a) corresponds to the Na atom and $I_0 = 2.5\times10^{13}$ W/cm$^2$, $\Delta t_p = 5$ fs; figure (b) corresponds to the Mg$^+$ ion and $I_0 = 2.5\times10^{13}$ W/cm$^2$, $\Delta t_p = 5$ fs. The black dashed line in both figures shows the time envelope of the laser field intensity.

# Generation of Subfemtosecond Deep and Vacuum UV pulses via Two-Photon Rabi Oscillations in Alkali Atoms or Alkaline Earth Ions

## SUPPLEMENTAL MATERIAL

I.R. Khairulin[1,2,*], A.A. Romanov[1], A.A. Silaev[1], M.Yu. Ryabikin[1,2] and V.A. Antonov[1]

[1]A.V. Gaponov-Grekhov Institute of Applied Physics of the Russian Academy of Sciences, Nizhny Novgorod, Russia

[2]Lobachevsky State University of Nizhny Novgorod, Nizhny Novgorod, Russia

In this Supplemental Material, we describe the methods of numerical and analytical solution of the time-dependent Schrödinger equation (TDSE) used for the calculation of dipole acceleration of alkali atoms and alkaline earth ions induced by a femtosecond pulse of an intense linearly polarized laser field under the conditions of two-photon resonant excitation of an atom (ion). We also present the parameters of the models used and the criteria implied to determine the optimal conditions for generating a shortened third harmonic pulse. Finally, we present the extended results of the numerical and analytical solutions of the TDSE and a comparison between them.

## 1. Theoretical model and description of the method used for numerical solution of the TDSE

The direct numerical simulation of the interaction of Na atom and $Mg^+$ ion with an intense linearly polarized laser pulse is performed on the basis of the solution of the three-dimensional TDSE within the approximation of a single active electron initially staying in the valence 3$s$ state:

$$i\frac{\partial \psi}{\partial t} = \left[-\frac{1}{2}\nabla^2 + zE_L(t) + V(r)\right]\psi. \quad \text{(S1)}$$

Here, $\psi(\mathbf{r},t)$ is the electron wave function and $E_L(t)$ is the laser electric field directed along $z$ axis, which is calculated as a time derivative of the laser vector potential projection on the $z$-axis $A(t)$,

$$E_L(t) = -\frac{1}{c}\frac{\partial A}{\partial t}, \quad \text{(S2)}$$

where

$$A(t) = -\frac{c}{\Omega}E_0 \exp\left(-2\ln(2)t^2/\Delta t_p^2\right)\sin(\Omega t), \quad \text{(S3)}$$

$c$ is the speed of light in vacuum, $E_0$ is the peak amplitude of the laser electric field, $\Omega$ is the laser carrier frequency, and $\Delta t_p$ is the laser pulse duration (the full width at half-maximum of the Gaussian intensity envelope). The definition (S2), (S3) ensures that the constant component of the electric field $E_L(t)$ is equal to zero for arbitrary parameters of the laser pulse. Hereinafter, the atomic system of units is used.

For the Na atom, we used the analytical expression for the parent ion potential $V(r)$ obtained in [1]. In turn, for the $Mg^+$ ion we calculated the parent ion potential $V(r)$ based on the numerical solution of the stationary Kohn-Sham equations using the LB94 exchange-correlation potential [2]. The used potentials ensure that the energies of the resonant bound states (see Fig. 1 in the main text) and the dipole matrix elements between them are close to the corresponding experimental data for Na and $Mg^+$ from [3] (see Section 6, Tables 1 and 2,3, respectively). We used the evolution in imaginary time to find the wavefunctions and energies of the stationary bound states in the potential $V(r)$. The initial condition for TDSE is the $|3s\rangle$ state in the potential $V(r)$.

The induced dipole acceleration $\ddot{d}_{ind}(t)$ (the subscript "ind" stands for induced), which is the second derivative of the dipole moment $d_{ind}(t) = -\langle \psi | z | \psi \rangle$, acquired by a quantum system under the action of an external electric field (S2), is found using the Ehrenfest theorem:

$$\ddot{d}_{ind}(t) = E_L(t) + \langle \psi | \frac{\partial V}{\partial z} | \psi \rangle. \tag{S4}$$

To obtain the complex amplitude of the third harmonic of the induced dipole acceleration, we calculate the Hilbert transform, $H(\ldots)$, of the dipole acceleration at the third harmonic of the laser frequency:

$$\tilde{\ddot{d}}_{3H}(t) = H\left( 2\mathrm{Re}\left\{ F^{-1}\left[ S_f(\omega) F\left( T_f(t) \ddot{d}_{ind}(t) \right) \right] \right\} \right)(t), \tag{S5}$$

where $F(\ldots)$ and $F^{-1}(\ldots)$ are direct and inverse fast Fourier transforms, respectively, $\mathrm{Re}(\ldots)$ is the real part; $T_f(t)$ is the time mask that allows to match the initial and final values of the function $\ddot{d}_{ind}(t)$ on the time grid $t_{min} \leq t \leq t_{max}$ for correct calculation of the fast Fourier transform:

$$T_f(t) = \begin{cases} \sin^2\left( 0.5\pi\Omega(t - t_{min})/(6\pi) \right), & t_{min} \leq t < t_{min} + 6\pi/\Omega, \\ 1, & t_{min} + 6\pi/\Omega \leq t < t_{max} - 6\pi/\Omega, \\ \cos^2\left( 0.5\pi\Omega(t - t_{max} + 6\pi/\Omega)/(6\pi) \right), & t_{max} - 6\pi/\Omega \leq t \leq t_{max}; \end{cases} \tag{S6}$$

$S_f(\omega)$ is the spectral filter that selects the frequency components of the dipole acceleration in the vicinity of the third harmonic frequency:

$$S_f(\omega) = \begin{cases} \sin^2\left( 0.5\pi(\omega/\Omega - 2)/0.25 \right), & 2 \leq \omega/\Omega < 2 + 0.25, \\ 1, & 2 + 0.25 \leq \omega/\Omega < 4 - 0.25, \\ \cos^2\left( 0.5\pi(\omega/\Omega - 4 + 0.25)/0.25 \right), & 4 - 0.25 \leq \omega/\Omega < 4, \\ 0, & \omega/\Omega < 2 \text{ and } \omega/\Omega \geq 4. \end{cases} \tag{S7}$$

To estimate the degree of ionization, we calculate the norm of the wave function inside the sphere of radius $r_0$:

$$n(t) = \int_{r < r_0} |\psi(\mathbf{r}, t)|^2 \, d\mathbf{r}. \tag{S8}$$

The value $r_0$ is chosen large enough to assume that $1 - n(t)$ estimates the ionization degree of the target. In our calculations we consider $r_0 = 20$ a.u. for both Na and Mg$^+$.

The TDSE (S1) is solved in spherical system of coordinates $(r, \theta, \varphi)$ using the expansion of the wavefunction in spherical harmonics [2, 4],

$$\psi(\mathbf{r}, t) = \frac{1}{r} \sum_{l=0}^{l_{max}} \Psi_l(r, t) Y_{l0}(\theta, \varphi), \tag{S9}$$

where $Y_{l0}(\theta, \phi) = \sqrt{(2l+1)/(4\pi)} P_l(\cos\theta)$, $l$ is the angular momentum quantun number, $P_l(\cos\theta)$ are Legendre polynomials, and $\theta$, $\varphi$ are the polar and azimuthal spherical angles, respectively, corresponding to the polar axis $z$. Due to the cylindrical symmetry of the problem and initial $s$ state, the magnetic quantum number $m$ is restricted to zero. Substituting (S9) in the TDSE, we obtain a system of equations for $\Psi_l$:

$$i\frac{\partial \Psi_l}{\partial t} = \left[-\frac{1}{2}\frac{\partial^2}{\partial r^2} + \frac{l(l+1)}{2r^2} + V(r)\right]\Psi_l + rE_L(t)\left(c_{l-1}\Psi_{l-1} + c_l\Psi_{l+1}\right), \tag{S10}$$

where $c_l = (l+1)[(2l+1)(2l+3)]^{-1/2}$. The system of equations (S10) is solved based on split-step and Crank-Nicolson methods using the finite difference discretization of the radial coordinate with the Numerov approximation for the second-order derivative [2]. The maximum orbital momentum in the expansion of wave functions in spherical harmonics is $l_{max} = 128$, and the numerical time step is $\Delta t = 0.02$ a.u. We use a nonuniform spatial grid, which has a higher density of nodes near the nucleus. The radial nodes of the spatial grid are specified as

$$r_k = k\Delta r + (\delta r/\Delta r - 1)r_\alpha \tanh(k\Delta r/r_\alpha), \tag{S11}$$

where $k$ is an integer, $\delta r = 10^{-4}$ a.u. is the tiny step of the radial grid near the nucleus, $\Delta r = 0.1$ a.u. is the radial step for large distances; $r_\alpha = 20$ a.u. is the scale for changing spatial step from $\delta r$ to $\Delta r$. The size of the radial grid was limited by $\max(\Delta r_{abs} + 3r_{osc}, 100 \text{ a.u.})$, where $r_{osc}$ is the maximum oscillatory radius of a free electron in the laser field and $\Delta r_{abs}$ is the width of the absorbing layer. The absorbing layer consists of a multi-hump imaginary potential [4] with the total width of $\Delta r_{abs} = 50$ a.u.

## 2. Derivation of the analytical solution

Let us consider a four-level model of the Na atom or Mg$^+$ ion, including the resonance states $|1\rangle - |4\rangle$, see Fig. 1 from the main text of the article. In this case, the wave function of the active electron of the atom (ion) can be represented as

$$|\psi\rangle = \sum_{k=1}^{4} a_k(t)|k\rangle, \tag{S12}$$

where $a_k(t)$ are the probability amplitudes of the states taken into account.

We will assume that an atom (ion) is irradiated by an intense laser pulse, the electric field of which has the form

$$E_L(t) = \tilde{E}_L(t)\cos(\Omega t) = \tilde{E}_L(t)\frac{e^{i\Omega t} + e^{-i\Omega t}}{2}, \tag{S13}$$

where $\Omega$ is the carrier frequency of the laser field, and $\tilde{E}_L(t)$ is its envelope, which we will assume to be a real function that slowly changes on the scale of the field oscillation period $T = 2\pi/\Omega$; in addition, $\tilde{E}_L(t \to -\infty) = 0$. Using definition (S13) instead of (S2) and (S3) allows us to obtain the analytical solution presented below. At the same time, for a Gaussian field envelope, $\tilde{E}_L(t) = E_0 \exp(-2\ln(2)t^2/\Delta t_p^2)$, the expressions (S2), (S3), and (S13) are practically equivalent for all combinations of field duration and frequency considered in the article, see Fig. S1.

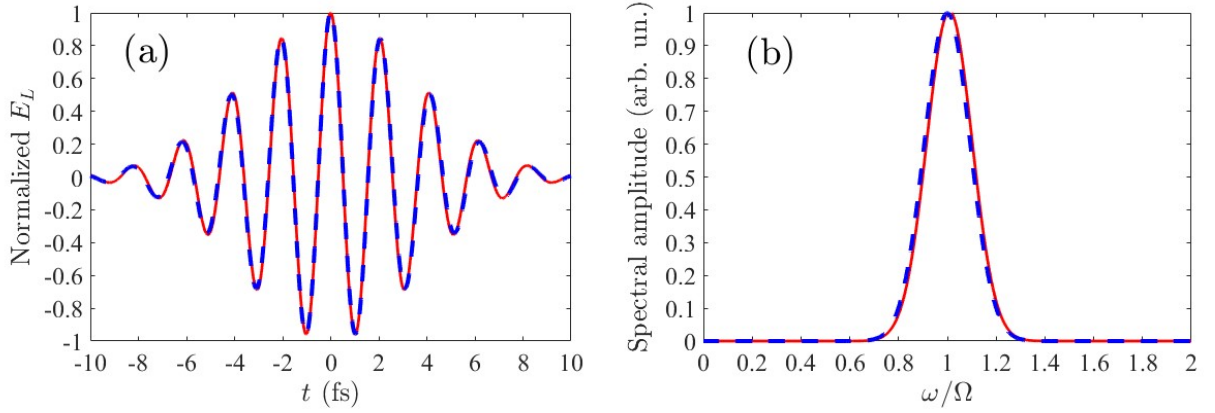

Fig. S1. Comparison of (a) the time profile and (b) amplitude spectra of the laser field calculated using formulas (S2), (S3) (red solid line), and (S13) (blue dotted line) for a laser pulse with a duration of $\Delta t_p = 5$ fs and photon energy of $\hbar\Omega = 1.99$ eV.

Substituting the expansion (S12) into (S1), we obtain a system of equations for the amplitudes $a_k(t)$:

$$\frac{da_1}{dt} = -i\mathrm{E}_1 a_1 - id_{12}E_L(t)a_2,$$

$$\frac{da_2}{dt} = -i\mathrm{E}_2 a_2 - id_{12}E_L(t)a_1 - id_{23}E_L(t)a_3 - id_{24}E_L(t)a_4,$$

$$\frac{da_3}{dt} = -i\mathrm{E}_3 a_3 - id_{23}E_L(t)a_2,$$

$$\frac{da_4}{dt} = -i\mathrm{E}_4 a_4 - id_{24}E_L(t)a_2.$$
(S14)

Here $\mathrm{E}_k$ is the energy of the state $|k\rangle$ ($k = 1,2,3,4$), $d_{kl}$ ($kl = 12, 23, 24$) is the dipole moment of transition $|k\rangle \leftrightarrow |l\rangle$. We will look for a solution to the system of equations (S14) in the form

$$a_k(t) = \sum_{n=-\infty}^{\infty} a_{k,n}(t) e^{-in\Omega t}, \quad k = 1, 2, 3, 4.$$
(S15)

Substituting (S15) into (S14) and equating the factors at the similar exponents, $\exp(-in\Omega t)$, we obtain a system of equations for $a_{k,n}(t)$ (the quasi-energy representation, see [5]):

$$\frac{da_{1,n}}{dt} = -i(\mathrm{E}_1 - n\Omega)a_{1,n} - i\frac{d_{12}\tilde{E}_L}{2}a_{2,n+1} - i\frac{d_{12}\tilde{E}_L}{2}a_{2,n-1},$$

$$\frac{da_{2,n}}{dt} = -i(\mathrm{E}_2 - n\Omega)a_{2,n} - i\frac{d_{12}\tilde{E}_L}{2}a_{1,n+1} - i\frac{d_{12}\tilde{E}_L}{2}a_{1,n-1}$$
$$- i\frac{d_{23}\tilde{E}_L}{2}a_{3,n+1} - i\frac{d_{23}\tilde{E}_L}{2}a_{3,n-1}$$
$$- i\frac{d_{24}\tilde{E}_L}{2}a_{4,n+1} - i\frac{d_{24}\tilde{E}_L}{2}a_{4,n-1},$$

$$\frac{da_{3,n}}{dt} = -i(\mathrm{E}_3 - n\Omega)a_{3,n} - i\frac{d_{23}\tilde{E}_L}{2}a_{2,n+1} - i\frac{d_{23}\tilde{E}_L}{2}a_{2,n-1},$$

$$\frac{da_{4,n}}{dt} = -i(\mathrm{E}_4 - n\Omega)a_{4,n} - i\frac{d_{24}\tilde{E}_L}{2}a_{2,n+1} - i\frac{d_{24}\tilde{E}_L}{2}a_{2,n-1}.$$
(S16)

Since in the Na atom or $Mg^+$ ion the frequencies of transitions $|1\rangle \leftrightarrow |2\rangle$, $|2\rangle \leftrightarrow |3\rangle$, and $|2\rangle \leftrightarrow |4\rangle$ are close (see Section 6, Tables 1, 2), then in the following, when deriving the

analytical solution, we will neglect the difference between them. In addition, we will assume that the carrier frequency of the laser field is exactly equal to the frequencies of these transitions. In this case, $E_1 = 0$, $E_2 = \Omega$, $E_3 = E_4 = 2\Omega$, and the system of equations (S16) takes the form

$$\frac{da_{1,n}}{dt} = -i(-n\Omega)a_{1,n} - i\frac{d_{12}\tilde{E}_L}{2}a_{2,n+1} - i\frac{d_{12}\tilde{E}_L}{2}a_{2,n-1},$$

$$\frac{da_{2,n}}{dt} = -i(1-n)\Omega a_{2,n} - i\frac{d_{12}\tilde{E}_L}{2}a_{1,n+1} - i\frac{d_{12}\tilde{E}_L}{2}a_{1,n-1}$$
$$- i\frac{d_{23}\tilde{E}_L}{2}a_{3,n+1} - i\frac{d_{23}\tilde{E}_L}{2}a_{3,n-1}$$
$$- i\frac{d_{24}\tilde{E}_L}{2}a_{4,n+1} - i\frac{d_{24}\tilde{E}_L}{2}a_{4,n-1},$$

$$\frac{da_{3,n}}{dt} = -i(2-n)\Omega a_{3,n} - i\frac{d_{23}\tilde{E}_L}{2}a_{2,n+1} - i\frac{d_{23}\tilde{E}_L}{2}a_{2,n-1},$$

$$\frac{da_{4,n}}{dt} = -i(2-n)\Omega a_{4,n} - i\frac{d_{24}\tilde{E}_L}{2}a_{2,n+1} - i\frac{d_{24}\tilde{E}_L}{2}a_{2,n-1}.$$

(S17)

We will also assume that at $t = -\infty$ (before interaction with the field) the active electron in the atom (ion) is in the ground state $|1\rangle$: $a_1(t=-\infty)=1$, $a_k(t=-\infty)=0$ for $k = 2,3,4$. The transition to the representation (S15) leads to ambiguity in the choice of initial conditions for $a_{k,n}$, which does not lead to other physical results. For simplicity, we will further use the initial conditions for $a_{k,n}$ in the form

$$a_{1,0}(t=-\infty) = 1,$$
$$a_{k,n}(t=-\infty) = 0, \ (k,n) \neq (1,0).$$

(S18)

A schematic representation of the system of equations (S17) with the initial conditions (S18) is shown in Fig. S2. Here, each amplitude $a_{k,n}$ corresponds to its own quasi-energy level, or quasi-level, the energy of which is determined by the frequency of the corresponding Fourier component of the probability amplitude and is $E_k - n\Omega$. In this case, under the action of the laser field, dipole transitions to other quasi-levels are possible, which differ in $n$ by one.

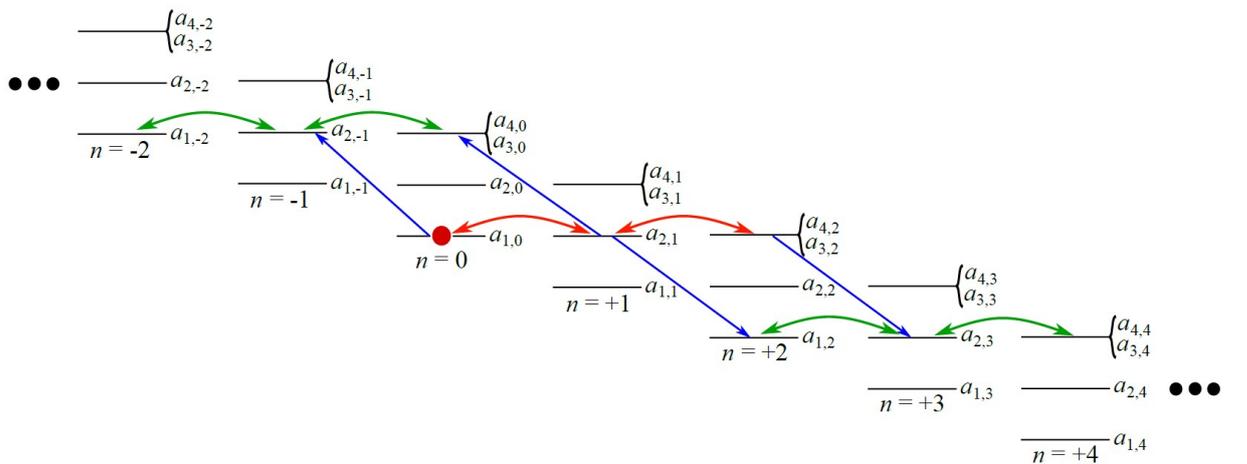

Fig. S2. Schematic representation of the system of equations (S17). The red circle denotes the initial condition (S18), the arrows denote the transitions that are taken into account in the analytical solution, and the color of the arrows denotes their type. Red arrows denote resonant transitions between resonantly populated quasi-levels; blue arrows denote transitions to non-resonant quasi-levels taken into account, the excitation of which leads to the generation of the third harmonic; green arrows denote resonant transitions between non-resonant quasi-levels.

In addition to the above assumptions, we will assume that the maximum Rabi frequency of the laser field at the transitions under consideration is significantly lower than its carrier frequency, i.e. the inequality is satisfied

$$\frac{d_{ij}E_0}{\Omega} \ll 1, \tag{S19}$$

where $E_0$ is the peak amplitude of the laser pulse, and $d_{ij}$ denotes any of the three dipole moments $d_{12}$, $d_{23}$, and $d_{24}$. In this case, from the entire infinite system of quasi-energy levels, only four quasi-levels with amplitudes $a_{1,0}$, $a_{2,1}$ and $a_{3,2}$, $a_{4,2}$ are most efficiently excited. The amplitudes of the remaining quasi-energy levels (Fourier components of the probability amplitudes of stationary states, see (S15)) are small, since their excitation is non-resonant (see Fig. S2). However, the excitation of resonant quasi-levels with amplitudes $a_{1,0}$, $a_{2,1}$ and $a_{3,2}$, $a_{4,2}$ does not lead to a dipole response of the atom at the frequency $3\Omega$. This is directly evident from the equation for the induced dipole moment of the atom $d_{ind}$ expressed through the amplitudes $a_{k,n}$:

$$d_{ind} = \sum_{k,s=1}^{4} d_{sk} \sum_{n,m=-\infty}^{\infty} a_{k,n}^* a_{s,m} \exp[-i(m-n)\Omega t],$$

$$\tilde{d}_{3H} = \sum_{n=-\infty}^{\infty} \sum_{k,s=1}^{4} d_{sk} a_{k,n}^* a_{s,n+3}, \tag{S20}$$

where $\tilde{d}_{3H}$ is the complex amplitude of the third harmonic in the spectrum of the induced dipole moment. Therefore, to describe the third harmonic in the dipole response of an atom (ion), it is necessary to take into account the excitation of non-resonant quasi-energy levels. When condition (S19) is satisfied, the greatest contribution to the response at the third harmonic frequency will be made by the quasi-levels closest to the resonant ones, which are divided into two groups, $a_{3,0}$, $a_{4,0}$, $a_{2,-1}$, $a_{1,-2}$ and $a_{1,2}$, $a_{2,3}$, $a_{3,4}$, $a_{4,4}$, in each of which the energies of the quasi-levels are equal to each other (see levels connected by green arrows in Fig. S2):

$$\tilde{d}_{3H} \simeq d_{12}a_{1,-2}^* a_{2,1} + d_{23}a_{2,-1}^* a_{3,2} + d_{24}a_{2,-1}^* a_{4,2} + + d_{23}a_{3,0}^* a_{2,3} +$$
$$+ d_{24}a_{4,0}^* a_{2,3} + d_{12}a_{1,0}^* a_{2,3} + d_{23}a_{2,1}^* a_{3,4} + d_{24}a_{2,1}^* a_{4,4}. \tag{S21}$$

In this case, since non-resonant quasi-levels are weakly excited, their influence on the excitation amplitudes of resonant quasi-levels is insignificant. Thus, the calculation of the response of an atom or ion at the frequency of the third harmonic of the laser field is reduced to the successive solution of three tasks. The first of these is describing the dynamics of resonant excitation of an atom (ion), which is reduced to solving the equations for the amplitudes $a_{k,n}$ of resonant quasi-levels:

$$\begin{cases} \dfrac{da_{1,0}}{dt} = -i\dfrac{d_{12}\tilde{E}_L}{2}a_{2,1}, \\ \dfrac{da_{2,1}}{dt} = -i\dfrac{d_{12}\tilde{E}_L}{2}a_{1,0} - i\dfrac{\tilde{E}_L}{2}F_2, \\ \dfrac{dF_2}{dt} = -i\dfrac{(d_{23}^2 + d_{24}^2)\tilde{E}_L}{2}a_{2,1}, \end{cases} \tag{S22}$$

where $F_2 = d_{23}a_{3,2} + d_{24}a_{4,2}$. The second task relates to the description of the excitation dynamics of non-resonant quasi-levels of the first group with the dependences of the amplitudes $a_{1,0}$, $a_{2,1}$ and $F_2$ on time obtained from (S22):

$$\begin{cases} \dfrac{da_{1,2}}{dt} = i2\Omega a_{1,2} - i\dfrac{d_{12}\tilde{E}_L}{2}a_{2,3} - i\dfrac{d_{12}\tilde{E}_L}{2}a_{2,1}, \\ \dfrac{da_{2,3}}{dt} = i2\Omega a_{2,3} - i\dfrac{d_{12}\tilde{E}_L}{2}a_{1,2} - i\dfrac{\tilde{E}_L}{2}F_4 - i\dfrac{\tilde{E}_L}{2}F_2, \\ \dfrac{dF_4}{dt} = i2\Omega F_4 - i\dfrac{(d_{23}^2 + d_{24}^2)\tilde{E}_L}{2}a_{2,3}, \end{cases} \quad (S23)$$

where $F_4 = d_{23}a_{3,4} + d_{24}a_{4,4}$. In turn, the third task relates to the description of the excitation dynamics of non-resonant quasi-levels of the second group:

$$\begin{cases} \dfrac{dF_0}{dt} = -i2\Omega F_0 - i\dfrac{(d_{23}^2 + d_{24}^2)\tilde{E}_L}{2}a_{2,-1} - i\dfrac{(d_{23}^2 + d_{24}^2)\tilde{E}_L}{2}a_{2,1}, \\ \dfrac{da_{2,-1}}{dt} = -i2\Omega a_{2,-1} - i\dfrac{d_{12}\tilde{E}_L}{2}a_{1,-2} - i\dfrac{d_{12}\tilde{E}_L}{2}a_{1,0} - i\dfrac{\tilde{E}_L}{2}F_0, \\ \dfrac{da_{1,-2}}{dt} = -i2\Omega a_{1,-2} - i\dfrac{d_{12}\tilde{E}_L}{2}a_{2,-1}, \end{cases} \quad (S241)$$

where $F_0 = d_{23}a_{3,0} + d_{24}a_{4,0}$. Note that due to the degeneracy of states $|3\rangle$ and $|4\rangle$, the products of the Fourier components of their amplitudes by the dipole moments of the corresponding transitions enter into Eqs. (S22)–(S24) in the form of linear combinations

$$F_n = d_{23}a_{3,n} + d_{24}a_{4,n}, \quad n = 0, 2, 4, \quad (S25)$$

the appearance of which means that the four-level model of an atom or ion under consideration is effectively a three-level model.

Let us start with solving the system of equations (S22). Instead of time $t$, we will use the local area of the laser pulse as the integration variable:

$$\xi = \int_{-\infty}^{t} \tilde{E}_L(t')dt'. \quad (S26)$$

Differentiating the equation for $a_{2,1}$ with respect to $\xi$ and taking into account the equations for $a_{1,0}$ and $F_2$, we obtain the equation of the harmonic oscillator with respect to $a_{2,1}$:

$$\dfrac{d^2 a_{2,1}}{d\xi^2} + D^2 a_{2,1} = 0, \quad (S27)$$

where

$$D = \sqrt{d_{12}^2 + d_{23}^2 + d_{24}^2}/2 \quad (S28)$$

has the meaning of the effective dipole moment of the system. In this case, the initial conditions for Eq. (S27) are

$$\begin{aligned} a_{2,1}\big|_{\xi=0} &= 0, \\ \dfrac{da_{2,1}}{d\xi}\bigg|_{\xi=0} &= -i\dfrac{d_{12}}{2}. \end{aligned} \quad (S29)$$

The solution of Eq. (S27) with initial conditions (S29) has the form

$$a_{2,1}(t) = -i\dfrac{d_{12}}{2D}\sin[D\xi(t)]. \quad (S30)$$

Next, substituting the solution (S30) into the equations for $a_{1,0}$ and $F_2$, taking into account the initial conditions (S18), we can obtain solutions for $a_{1,0}$ and $F_2$:

$$a_{1,0}(t) = 1 - \frac{d_{12}^2}{2D^2} \sin^2[D\xi(t)/2],$$

$$F_2(t) = -\frac{d_{12}(d_{23}^2 + d_{24}^2)}{2D^2} \sin^2[D\xi(t)/2],$$

(S31)

and taking into account the definition of $F_2$ (S24), we can obtain expressions for $a_{3,2}$ and $a_{4,2}$:

$$a_{3,2}(t) = -\frac{d_{12}d_{23}}{2D^2} \sin^2[D\xi(t)/2],$$

$$a_{4,2}(t) = -\frac{d_{12}d_{24}}{2D^2} \sin^2[D\xi(t)/2].$$

(S32)

Since the amplitudes of the remaining quasi-energy levels are significantly smaller than the amplitudes $a_{1,0}$, $a_{2,1}$, $a_{3,2}$, and $a_{4,2}$, solutions (S30), (S31), and (S32) describe the main dynamics of the change in populations $n_i(t), i = \{1,2,3,4\}$, of the $|1\rangle - |4\rangle$ states taken into account:

$$n_1(t) \simeq |a_{1,0}(t)|^2 = \left\{1 - \frac{d_{12}^2}{2D^2} \sin^2[D\xi(t)/2]\right\}^2,$$

$$n_2(t) \simeq |a_{2,1}(t)|^2 = \frac{d_{12}^2}{4D^2} \sin^2[D\xi(t)],$$

$$n_3(t) \simeq |a_{3,2}(t)|^2 = \frac{d_{12}^2 d_{23}^2}{4D^4} \sin^4[D\xi(t)/2],$$

$$n_4(t) \simeq |a_{4,2}(t)|^2 = \frac{d_{12}^2 d_{24}^2}{4D^4} \sin^4[D\xi(t)/2].$$

(S33)

As follows from these solutions, under the action of a resonant laser pulse, the populations of the states under consideration oscillate in time with an increase in the instantaneous value of the area under the laser pulse $D\xi(t)$.

Let us now consider the system of equations (S23) describing the excitation dynamics of non-resonant quasi-levels of the first group. We will look for its solution in the form

$$a_{1,2} = \hat{a}_{1,2} e^{i2\Omega t},$$

$$a_{2,3} = \hat{a}_{2,3} e^{i2\Omega t},$$

$$F_4 = \hat{F}_4 e^{i2\Omega t}.$$

(S34)

By introducing the change of variable (S26), the system of equations (S23) can be rewritten as

$$\begin{cases} \dfrac{d\hat{a}_{1,2}}{d\xi} = -i\dfrac{d_{12}}{2}\hat{a}_{2,3} - i\dfrac{d_{12}e^{-i2\Omega t}}{2}a_{2,1}, \\ \dfrac{d\hat{a}_{2,3}}{d\xi} = -i\dfrac{d_{12}}{2}\hat{a}_{1,2} - i\dfrac{1}{2}\hat{F}_4 - i\dfrac{e^{-i2\Omega t}}{2}F_2, \\ \dfrac{d\hat{F}_4}{d\xi} = -i\dfrac{(d_{23}^2 + d_{24}^2)}{2}\hat{a}_{2,3}, \end{cases}$$

(S35)

where the initial conditions for the functions $\hat{a}_{1,2}$, $\hat{a}_{2,3}$, and $\hat{F}_4$ have the form

$$\hat{a}_{1,2}\big|_{\xi=0} = 0,$$

$$\hat{a}_{2,3}\big|_{\xi=0} = 0,$$

$$\hat{F}_4\big|_{\xi=0} = 0.$$

(S36)

Differentiating the second equation in (S35) with respect to $\xi$ and substituting the derivatives from the first and third equations, we obtain the equation of a harmonic oscillator with non-zero external force:

$$\frac{d^2 \hat{a}_{2,3}}{d\xi^2} = -D^2 \hat{a}_{2,3} - \frac{d_{12}^2}{4} a_{2,1} e^{-i2\Omega t} - \frac{i}{2}\frac{d}{d\xi}\left(F_2 e^{-i2\Omega t}\right). \tag{S37}$$

Using the method of variation of arbitrary constants, taking into account the substitution (S34), we obtain the following solution for $a_{2,3}$:

$$a_{2,3}(t) = -\frac{d_{12}^2}{4D}\int_{-\infty}^{t} \tilde{E}_L(t') a_{2,1}(t') \sin\left[D(\xi(t) - \xi(t'))\right] e^{i2\Omega(t-t')} dt' + \\ -\frac{i}{2}\int_{-\infty}^{t} \tilde{E}_L(t') F_2(t') \cos\left[D(\xi(t) - \xi(t'))\right] e^{i2\Omega(t-t')} dt'. \tag{S38}$$

If condition (S19) is satisfied, the approximate solution for the integrals in (S38) can be found analytically via integration by parts. The corresponding expression, accurate to the second order in the parameter $d_{ij} E_0/\Omega$, has the form

$$a_{2,3}(t) \simeq -\frac{(d_{23}^2 + d_{24}^2 - d_{12}^2)\tilde{E}_L^2}{16\Omega^2} a_{2,1} + \frac{1}{4\Omega}\tilde{E}_L F_2 - \frac{iF_2}{8\Omega^2}\frac{d\tilde{E}_L}{dt}, \tag{S39}$$

where $a_{2,1}$ and $F_2$ are determined by Eqs. (S30) and (S31). Next, substituting solution (S39) into Eq. (S23), for $F_4$ and $a_{1,2}$ we obtain with an accuracy of up to the second order in the parameter $d_{ij} E_0/\Omega$:

$$F_4(t) \simeq \frac{(d_{23}^2 + d_{24}^2)}{16\Omega^2}\left(\tilde{E}_L^2 + \frac{1}{i4\Omega}\frac{d(\tilde{E}_L^2)}{dt}\right) F_2,$$

$$a_{1,2}(t) = \frac{d_{12}}{4\Omega}\left(\tilde{E}_L + \frac{1}{i2\Omega}\frac{d\tilde{E}_L}{dt}\right) a_{2,1} - \frac{d_{12}^2 \tilde{E}_L^2}{16\Omega^2} a_{1,0} - \frac{id_{12}}{64\Omega^3} F_2 \frac{d(\tilde{E}_L^2)}{dt}. \tag{S40}$$

Finally, let us consider the system of equations (S24). We will look for its solution in the form

$$F_0 = \hat{F}_0 e^{-i2\Omega t},$$
$$a_{2,-1} = \hat{a}_{2,-1} e^{-i2\Omega t}, \tag{S41}$$
$$a_{1,-2} = \hat{a}_{1,-2} e^{-i2\Omega t}.$$

Then, introducing the change of variable (S26), we obtain the following system of equations for $\hat{F}_0$, $\hat{a}_{2,-1}$, and $\hat{a}_{1,-2}$:

$$\begin{cases} \dfrac{d\hat{F}_0}{d\xi} = -i\dfrac{(d_{23}^2 + d_{24}^2)}{2}\hat{a}_{2,-1} - i\dfrac{(d_{23}^2 + d_{24}^2)}{2} a_{2,1} e^{i2\Omega t}, \\ \dfrac{d\hat{a}_{2,-1}}{d\xi} = -i\dfrac{d_{12}}{2}\hat{a}_{1,-2} - i\dfrac{1}{2}\hat{F}_0 - i\dfrac{d_{12}}{2} a_{1,0} e^{i2\Omega t}, \\ \dfrac{d\hat{a}_{1,-2}}{d\xi} = -i\dfrac{d_{12}}{2}\hat{a}_{2,-1}. \end{cases} \tag{S42}$$

Equations (S42) can be reduced to a system of equations for $\hat{a}_{2,-1}$ and a function $\Phi = d_{12}\hat{a}_{1,-2} + \hat{F}_0$:

$$\begin{cases} \dfrac{d\Phi}{d\xi} = -i2D^2 \hat{a}_{2,-1} - i\dfrac{(d_{23}^2 + d_{24}^2)}{2} a_{2,1} e^{i2\Omega t}, \\ \dfrac{d\hat{a}_{2,-1}}{d\xi} = -\dfrac{i}{2}\Phi - i\dfrac{d_{12}}{2} a_{1,0} e^{i2\Omega t}. \end{cases} \tag{S43}$$

Expressing $\hat{a}_{2,-1}$ from the first equation in (S43) and substituting it into the second equation, we obtain

$$\frac{d^2\Phi}{d\xi^2} + D^2\Phi = -d_{12}D^2 a_{1,0} e^{i2\Omega t} - i\frac{(d_{23}^2 + d_{24}^2)}{2}\frac{d}{d\xi}\left(a_{2,1} e^{i2\Omega t}\right). \tag{S44}$$

The solution of Eq. (S44) can be found by the method of variation of arbitrary constants similarly to Eq. (S37). It is important to note that, due to the construction of the solution by this method, we have

$$\frac{d\Phi}{d\xi} = iDC_1(\xi)e^{iD\xi} - iDC_2(\xi)e^{-iD\xi}. \tag{S45}$$

From the first equation of system (S43), taking into account (S45), we have

$$\hat{a}_{2,-1} = -\frac{(d_{23}^2 + d_{24}^2)}{4D^2} a_{2,1} e^{i2\Omega t} - \frac{1}{2D}\left[C_1(\xi)e^{iD\xi} - C_2(\xi)e^{-iD\xi}\right]. \tag{S46}$$

Calculating $C_1(\xi)$ and $C_2(\xi)$ and substituting them into (S46), taking into account (S41), we obtain the following expression for $a_{2,-1}(t)$:

$$a_{2,-1}(t) = -\frac{id_{12}}{2}\int_{-\infty}^{t}\tilde{E}_L a_{1,0} \cos[D(\xi(t) - \xi(t'))] e^{-i2\Omega(t-t')} dt' - \\ -\frac{(d_{23}^2 + d_{24}^2)}{4D}\int_{-\infty}^{t}\tilde{E}_L a_{2,1} \sin[D(\xi(t) - \xi(t'))] e^{-i2\Omega(t-t')} dt'. \tag{S47}$$

Calculating the integrals in (S47) with an accuracy of up to the second order in the parameter $d_{ij}E_0/\Omega$, we obtain

$$a_{2,-1}(t) \simeq \frac{(d_{23}^2 + d_{24}^2 - d_{12}^2)\tilde{E}_L^2}{16\Omega^2} a_{2,1} - \frac{d_{12}}{4\Omega}\tilde{E}_L a_{1,0} - \frac{id_{12}}{8\Omega^2} a_{1,0} \frac{d\tilde{E}_L}{dt}. \tag{S48}$$

Next, substituting solution (S48) into Eqs. (S24) for $F_0$ and $a_{1,-2}$, we obtain, with an accuracy of up to the second order in parameter $d_{ij}E_0/\Omega$:

$$F_0(t) \simeq i\frac{d_{12}(d_{23}^2 + d_{24}^2)}{64\Omega^3}\frac{d(\tilde{E}_L^2)}{dt} a_{1,0} - \frac{(d_{23}^2 + d_{24}^2)}{4\Omega}\left(\tilde{E}_L - \frac{1}{i2\Omega}\frac{d\tilde{E}_L}{dt}\right) a_{2,1} - \frac{(d_{23}^2 + d_{24}^2)\tilde{E}_L^2}{16\Omega^2} F_2, \\ a_{1,-2}(t) \simeq \frac{d_{12}^2}{16\Omega^2}\left(\tilde{E}_L^2 - \frac{1}{i4\Omega}\frac{d(\tilde{E}_L^2)}{dt}\right) a_{1,0}. \tag{S49}$$

Taking into account the notation (S25), the expression (S21) for the slowly varying amplitude of the dipole moment of an atom at the frequency of the third harmonic of the laser field, $3\Omega$, can be written as

$$\tilde{d}_{3H} \simeq d_{12} a_{1,-2}^* a_{2,1} + d_{12} a_{1,0}^* a_{2,3} + F_2 a_{2,-1}^* + F_0^* a_{2,3} + F_4 a_{2,1}^*, \tag{S50}$$

where different terms correspond to different radiation pathways of the third harmonic.

Using Eqs. (S39), (S40), (S48), and (S49), up to the second order in the parameter $d_{ij}E_0/\Omega$, the terms in (S50) can be rewritten as follows:

$$d_{12}a^*_{1,-2}a_{2,1} = \frac{d_{12}^2}{16\Omega^2}\left(\tilde{E}_L^2 + \frac{1}{i4\Omega}\frac{d(\tilde{E}_L^2)}{dt}\right)d_{12}a^*_{1,0}a_{2,1},$$

$$d_{12}a^*_{1,0}a_{2,3} = \frac{d_{12}}{4\Omega}\left(\tilde{E}_L + \frac{1}{i2\Omega}\frac{d\tilde{E}_L}{dt}\right)F_2 a^*_{1,0} - \frac{(d_{23}^2 + d_{24}^2 - d_{12}^2)\tilde{E}_L^2}{16\Omega^2}d_{12}a_{2,1}a^*_{1,0},$$

$$F_2 a^*_{2,-1} = \frac{(d_{23}^2 + d_{24}^2 - d_{12}^2)\tilde{E}_L^2}{16\Omega^2}F_2 a^*_{2,1} - \frac{d_{12}}{4\Omega}\left(\tilde{E}_L + \frac{1}{i2\Omega}\frac{d\tilde{E}_L}{dt}\right)F_2 a^*_{1,0}, \quad \text{(S51)}$$

$$a_{2,3}F_0^* = -\frac{d_{23}^2 + d_{24}^2}{16\Omega^2}\left(\tilde{E}_L + \frac{1}{i2\Omega}\frac{d\tilde{E}_L}{dt}\right)^2 F_2 a^*_{2,1},$$

$$a^*_{2,1}F_4 = \frac{d_{23}^2 + d_{24}^2}{16\Omega^2}\left(\tilde{E}_L^2 + \frac{1}{i4\Omega}\frac{d(\tilde{E}_L^2)}{dt}\right)F_2 a^*_{2,1}.$$

In a laser field with a slowly varying amplitude we have $\left|d\tilde{E}_L/dt\right| \ll \Omega\tilde{E}_L$. Accordingly, in (S51) the coefficients with the field derivatives with respect to time are small. Neglecting them we obtain $F_0^* a_{2,3} + F_4 a^*_{2,1} = 0$, and the atomic dipole moment at the frequency of the third harmonic is determined only by the first three terms in Eq. (S50). In addition, the terms linear in the amplitude of the field and its derivative in the expressions for $d_{12}a^*_{1,0}a_{2,3}$ and $F_2 a^*_{2,-1}$ (the second and third lines in (S51)) exactly compensate each other. Thus, the final expression for the slowly varying amplitude of the third harmonic in the spectrum of the induced dipole moment of the atom has the form

$$\tilde{d}_{3H} \simeq -\frac{(d_{23}^2 + d_{24}^2 - 2d_{12}^2)\tilde{E}_L^2}{16\Omega^2}\left[d_{12}a_{2,1}a^*_{1,0} - \frac{d_{23}^2 + d_{24}^2 - d_{12}^2}{d_{23}^2 + d_{24}^2 - 2d_{12}^2}\left(d_{23}a_{3,2}a^*_{2,1} + d_{24}a_{4,2}a^*_{2,1}\right)\right]. \quad \text{(S52)}$$

Equation (S52) allows us to write down the solution for the intensity of the third harmonic of the laser field, $I_3$, at the output of an optically thin medium, proportional to the square modulus of the complex amplitude the dipole acceleration at the third harmonic frequency, $\left|\tilde{\ddot{d}}_{3H}\right|^2 \simeq 9\Omega^2\left|\tilde{d}_{3H}\right|^2$, with an accuracy of up to a dimensional coefficient (see, for example, [6]). Substituting into (S52) the expressions for $a_{1,0}$ (S31), $a_{2,1}$ (S30), $a_{3,2}$ and $a_{4,2}$ (S32), calculating the square of the modulus and multiplying it by $9\Omega^2$, we obtain

$$I_3(t) = \alpha\left(\tilde{E}_L/E_0\right)^4 R(D\xi) \quad \text{(S53)}$$

and

$$\alpha = \left[\frac{9}{32}\cdot\frac{d_{23}^2 + d_{24}^2 - 2d_{12}^2}{D}\cdot(d_{12}E_0)^2\right]^2,$$

$$R(D\xi) = \sin^2(D\xi)\left[1 + \frac{2d_{12}^4 - (d_{23}^2 + d_{24}^2)^2}{2D^2(d_{23}^2 + d_{24}^2 - 2d_{12}^2)}\sin^2(D\xi/2)\right]^2, \quad \text{(S54)}$$

coinciding with Eqs. (2) and (3) from the main text of the article.

## 3. Analysis of optimal conditions for generating a shortened third harmonic pulse in a medium of Na atoms based on an analytical TDSE solution

The analytical solution (2), (3), and (S33) is valid for a laser pulse (S13) with an arbitrary envelope $\tilde{E}_L(t)$. Further, we will assume that the envelope has a Gaussian shape with a full width at half maximum of the field intensity $\Delta t_p$:

$$\tilde{E}_L(t) = E_0 \exp\left(-2\ln(2) t^2 / \Delta t_p^2\right). \tag{S55}$$

From the solution (2), (3), and (S33) it is evident that for a given shape of the laser field envelope, the time dependences of the third harmonic intensity and the populations of the resonant states of the atom (ion) are determined exclusively by the total area of the laser pulse $DS_p$, where $S_p = \xi(t \to \infty)$ is proportional to the product of the laser field amplitude and its duration, $E_0 \Delta t_p$. Thus, determining the optimal conditions for generating a short third harmonic pulse is reduced to finding the optimal value of the laser pulse peak intensity $I_0 \sim E_0^2$ for its fixed duration $\Delta t_p$. Without loss of generality, we will consider below the case of $\Delta t_p = 20$ fs.

Let us consider the results of the analytical solution of (2), (3), and (S33) for the Na atom (see Section 6, Table 1), the characteristics of which are calculated on the basis of the single-electron potential used to solve the TDSE from first principles (see Section 1 of these Supplemental Materials). Figure S3 shows the dependences of the compression coefficient $\beta$, defined as the ratio of the laser pulse duration $\Delta t_p$ to the duration of the generated third harmonic pulse $\Delta t_{3H}$, see Eq. (5) of the main text (left axis, blue solid line), and the intensity ratio of the two most intense bursts in the time dependence of the third harmonic intensity, $I_3^{(max1)}/I_3^{(max2)}$, $I_3^{(max1)} > I_3^{(max2)}$ (right axis, red solid line), on the laser pulse intensity $I_0$ for the Na atom. In addition, for a reference, Fig. S3 shows a black dotted line depicting the compression coefficient for a nonresonant medium with cubic nonlinearity equal to $\beta_{nonres} = \sqrt{3}$. As follows from Fig. S3, for a laser pulse with low intensity, $I_0 < 10^{11}$ W/cm$^2$ (which for a Na atom and a pulse of duration $\Delta t_p = 20$ fs corresponds to $DS_p < \pi$), the compression coefficient $\beta$ monotonically decreases with decreasing $I_0$, while the ratio $I_3^{(max1)}/I_3^{(max2)}$ monotonically increases. In this case, the time dependence of the third harmonic intensity is determined by the product $I_3 \sim \tilde{E}_L^4(t)\xi^2(t)$ and has the form of a single burst, the duration of which increases with decreasing $I_0$. In this case, the probability of the atom excitation is small, the populations of the excited states $|2\rangle$–$|4\rangle$ are close to zero, and the population of the ground state $|1\rangle$ is close to unity (see Eqs. (S33)). At a higher intensity of the laser field, $I_0 > 10^{11}$ W/cm$^2$, the dependence of the compression coefficient $\beta$ on $I_0$ becomes piecewise continuous, which is associated with the appearance in the time dependence of the third harmonic intensity of several maxima of comparable amplitude and, as a consequence, jumps in the found full width at half maximum of the third harmonic intensity.

In the limit of a high-intensity laser pulse, $I_0 > 10^{13}$ W/cm$^2$ (which for the Na atom and a pulse of duration $\Delta t_p = 20$ fs corresponds to $DS_p \gg \pi$), the ratio of peak intensities $I_3^{(max1)}/I_3^{(max2)}$ is close to unity, and the compression coefficient $\beta$ with increasing $I_0$ is localized in the vicinity of the value $\sqrt{2}$, equal to the ratio of the laser pulse duration $\Delta t_p$ to the duration of the function $\tilde{E}_L^4(t)$. In this case, the time dependence $I_3(t)$ is a sequence of short bursts with the envelope $\left(\tilde{E}_L/E_0\right)^4$ (see (2) and (3)), and multiple Rabi oscillations occur between the resonance states of the atom (see (S33)). However, comparison with the numerical solution of the TDSE from first principles shows that the described case is unrealizable in practice, since a high-intensity laser pulse induces rapid ionization of the atom. As a consequence, most of the bursts on the trailing edge of the generated third harmonic predicted by the analytical solution (2) and (3) are suppressed, and its envelope is shortened. In the limiting case, this allows one to form a single third harmonic pulse extremely short compared to the duration of the laser field (S55). However, due to the rapid, almost complete, ionization of the atom in this regime, the resonant interaction of the laser field with the atoms is suppressed, and the efficiency of generating such a third harmonic pulse, characterized by the ratio of the harmonic intensity to the laser intensity, is low.

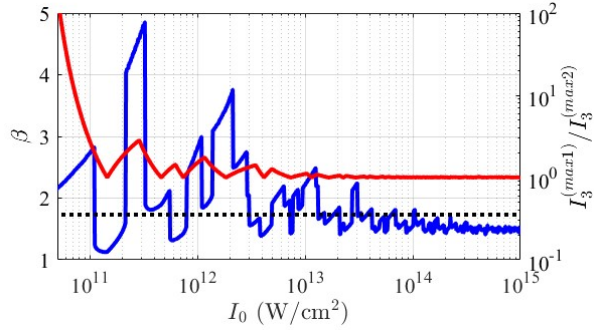

Fig. S3. Dependences of the compression coefficient $\beta$ (left axis, blue solid line) and the intensity ratio of the two most intense bursts (right axis, red solid line) on the laser pulse intensity, calculated based on the analytical solution (2) and (3) for the model Na atom (see Section 6, Table 1). The black dotted line (left axis) corresponds to the compression coefficient for a nonresonant medium with cubic nonlinearity. The laser pulse duration is assumed to be $\Delta t_p = 20$ fs.

At intermediate intensities, $10^{11}$ W/cm$^2$ $\ll I_0 < 10^{13}$ W/cm$^2$, there are ranges of $I_0$ values for which $\beta > \sqrt{3}$, i.e. the duration of the generated pulse in the time dependence of the third harmonic intensity is shorter than in the case of a non-resonant medium with cubic nonlinearity. In the former case, in all such regions except one, corresponding to $2.1 \times 10^{11}$ W/cm$^2 < I_0 < 3.2 \times 10^{11}$ W/cm$^2$, the ratio of peak intensities $I_3^{(max1)}/I_3^{(max2)}$ is close to unity. The generated third harmonic pulse then consists of two short and close to each other (so that their total duration is small compared to $\Delta t_p$) bursts of comparable amplitude. At the same time, in the range of $2.1 \times 10^{11}$ W/cm$^2 < I_0 < 3.2 \times 10^{11}$ W/cm$^2$, the values $I_3^{(max1)}/I_3^{(max2)} > 2$ and $\beta > 4$ are reached. In this case, the main burst in the time dependence of the third harmonic intensity is more than twice as intense as the side bursts, and its duration is shorter than 1/4 of the laser pulse duration. These conditions correspond to the formation of a single third-harmonic pulse of short (in comparison with both $\Delta t_p$ and the case of a nonresonant medium) duration. For the considered laser pulse duration, $\Delta t_p = 20$ fs, the optimal value of its intensity is $I_0 = 2.9 \times 10^{11}$ W/cm$^2$ and corresponds to the local maximum of the ratio of the intensities of the main and side bursts in the time dependence of the third harmonic radiation, $I_3^{(max1)}/I_3^{(max2)} \simeq 2.8$. The corresponding value of the total laser pulse area is $DS_p = 1.59\pi$, and the compression coefficient is $\beta = 4.6$.

Figure S4 shows the time dependences of the populations of the resonance states of the Na atom (Fig. S4(a)), the dependences of $R$ and $(\tilde{E}_L/E_0)^4$ on the normalized instantaneous area $\xi/S_p$ of the laser pulse (Fig. S4(b), red solid and black dashed lines, respectively), as well as the time dependences of the intensities of the third harmonic and the fundamental frequency laser pulse (Fig. S4(c), red solid and black dashed lines) and the dependence of $\xi/S_p$ on time (Fig. S4(c), blue dotted line). These dependences are obtained on the basis of the analytical solution (2), (3), and (S33) using the parameters of the Na atom from Table 1 at the found optimal values $\Delta t_p = 20$ fs and $I_0 = 2.9 \times 10^{11}$ W/cm$^2$ ($DS_p = 1.59\pi$).

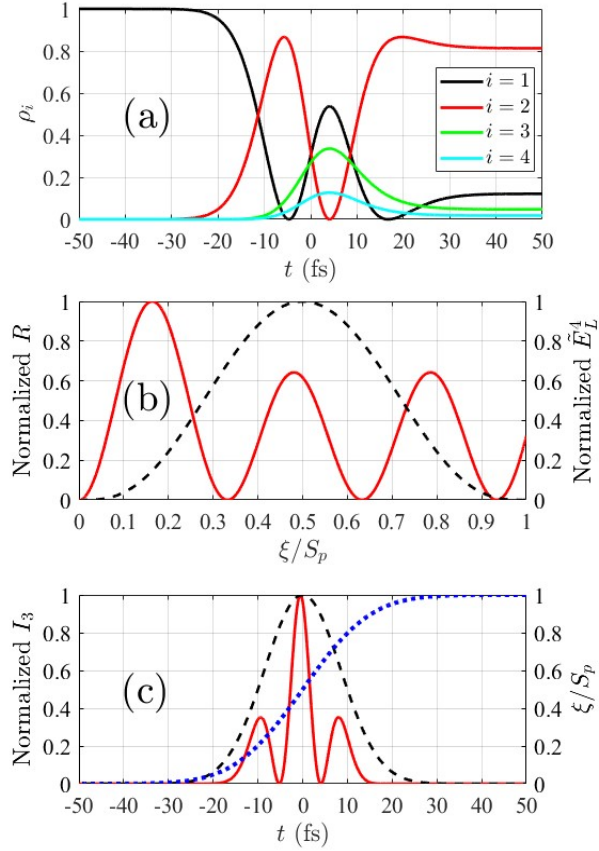

Fig. S4. (a) Time dependences of the populations of states $|i\rangle$, $i = 1,2,3,4$ of the model Na atom (see Section 6, Table 1). (b) Normalized dependences of the function $R$, see (3) and (S54) (left axis, red solid line), and the fourth power of the slowly varying amplitude of the Gaussian laser pulse (right axis, black dashed line) on the instantaneous value of the normalized area under the laser pulse. (c) Time dependences of the third harmonic intensity (left axis, red solid line) and the instantaneous area under the laser field pulse (right axis, blue dotted line) normalized to the maximum values. The figures are drawn based on the analytical solution (2), (3), and (S33) for the model Na atom; $\hbar\Omega = .99$ eV, $DS_p = 1.59\pi$. Figures (a) and (c) correspond to $\Delta t_p = 20$ fs and $I_0 = 2.9 \times 10^{11}$ W/cm$^2$.

It is seen that under optimal conditions, approximately one and a half Rabi oscillations occur between states $|1\rangle$ and $|2\rangle$, and states $|3\rangle$ and $|4\rangle$ are noticeably populated. In addition, the laser pulse envelope in this case covers three bursts in the dependence $R(\xi/S_p)$. The first and third bursts are then located at the leading and trailing edges of the laser pulse, where the value of $(\tilde{E}_L/E_0)^4$ is small, and the second burst is in the vicinity of the maximum $(\tilde{E}_L/E_0)^4$. As a result, see (2), in the time dependence of the third harmonic intensity, $I_3(t)$, one intense short pulse is formed, as well as two side bursts of equal intensity, which is approximately 2.8 times lower than the intensity of the main pulse. In this case, the full width at half-maximum of intensity of the formed pulse is $\Delta t_{3H} \approx 4.3$ fs, which is approximately 4.6 times shorter than the duration of the laser pulse $\Delta t_p$ and 2.7 times shorter than the similar duration in the case of a nonresonant medium with cubic nonlinearity, $\Delta t_{3H}^{(nonres)} = \Delta t_p / \sqrt{3}$.

## 4. Analysis of optimal conditions for generating a shortened third harmonic pulse in a medium of Mg$^+$ ions based on an analytical TDSE solution

Now let us consider the results of the analytical solution (2), (3), and (S33) for the Mg$^+$ ion (see Section 6, Table 2), the characteristics of which are calculated on the basis of the one-electron potential used to solve the TDSE from first principles (see Section 1), shown in Figs. S5 and S6. These figures are similar to Figs. S3 and S4 for Na atoms and explain the optimal conditions for generating a shortened third harmonic pulse in the medium of Mg$^+$ ions. As follows from these figures, the same features are preserved for the Mg$^+$ ion as for the Na atom; however, due to the difference in the dipole moments of the resonant transitions, the optimal values of the area and peak intensity of the laser pulse are different. Namely, for the model Mg$^+$ ion, the optimal value of the laser pulse area is $DS_p = 3.27\pi$, which for the pulse duration $\Delta t_p = 20$ fs corresponds to the field peak intensity $I_0 = 1.33 \times 10^{12}$ W/cm$^2$.

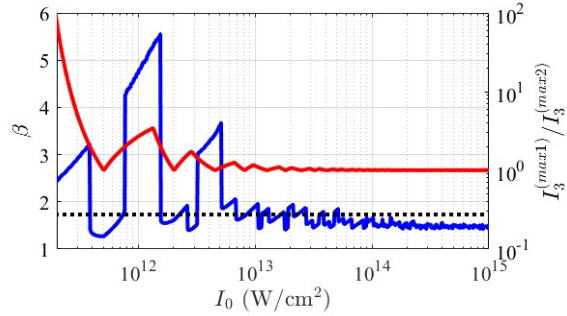

Fig. S5. Same as Fig. S3, but for the model Mg$^+$ ion (see text). The laser pulse duration is $\Delta t_p = 20$ fs.

The differences in (a) the transition dipole moments and (b) the optimal area of the laser pulse result in the corresponding differences in the time dependences of the populations of the resonance states, as well as the shape of the generated third harmonic pulse for Na atoms and Mg$^+$ ions (cf. Figs. S4 and S6). In particular, Rabi oscillations arise in the Mg$^+$ ion at the $|2\rangle \leftrightarrow |3\rangle$ and $|2\rangle \leftrightarrow |4\rangle$ transitions, which are absent in the Na atom. As a result, the oscillations in the time dependence of the intermediate state $|2\rangle$ population are faster, and the compression coefficient for Mg$^+$ ions under optimal conditions increases relative to Na atoms and reaches $\beta \simeq 5.4$.

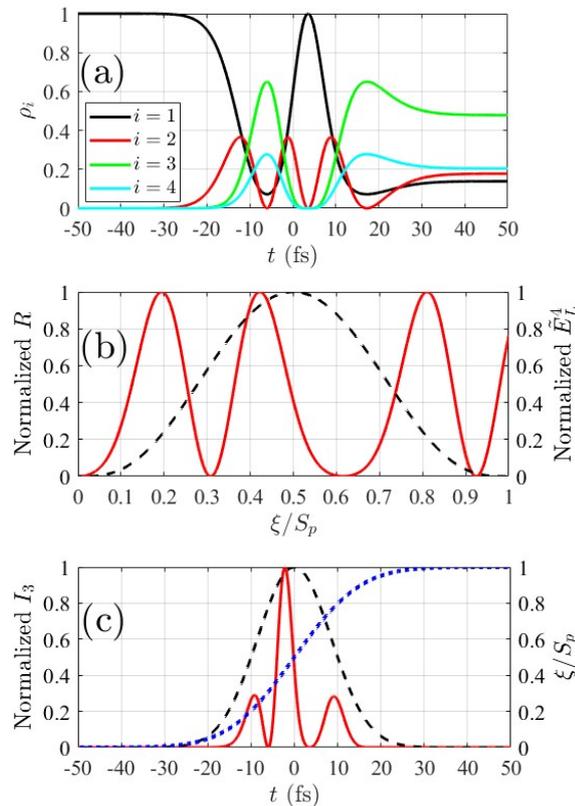

Fig. S6. Same as Fig. S4, but for the model Mg$^+$ ion (see text), $\hbar\Omega = 4.77$ eV, $DS_p = 3.27\pi$. Figures (a) and (c) correspond to $\Delta t_p = 20$ fs and $I_0 = 1.33 \times 10^{12}$ W/cm$^2$.

# 5. Results of the first-principles solution of the TDSE: efficiency of the third harmonic generation and stability of the solution to changes in laser parameters

The efficiency of third harmonic generation in resonance with a cascade two-photon transition of an atom or ion turns out to be significantly higher than far from it. Figure S7(a,b) shows the dependences of the peak intensity of the third harmonic generated by Na atoms (Fig. S7(a)) and $Mg^+$ ions (Fig. S7(b)) in a laser field with resonant (red circles connected by dashed lines) and nonresonant (black stars connected by dashed lines) carrier frequency on the laser pulse duration at the corresponding optimal intensity $I_0$ (see markers in Fig. 3(a)). Figure S7(c,d) shows the corresponding dependences of the electron wave function norm within a sphere of radius $r_0 = 20$ a.u. after the end of the laser pulse. The curves in Fig. S7(c,d) characterize the part of the electron wave packet remaining in the vicinity of the atomic or ionic core and allow one to estimate the degree of ionization of the atom or ion. As follows from Fig. S7, with a sufficiently long duration and moderate intensity of the laser field pulse, corresponding to a not too high degree of ionization (less than 20% for the Na atom and less than 50% for the $Mg^+$ ion), the peak intensity of the third harmonic in resonance with the two-photon transition turns out to be three orders of magnitude higher than in the nonresonant case. However, with decreasing duration and corresponding increase in intensity of the laser pulse, the degree of ionization of the atom or ion in the resonance case increases faster than far from resonance. As a result, the difference between the peak intensities in the resonance and nonresonance cases decreases.

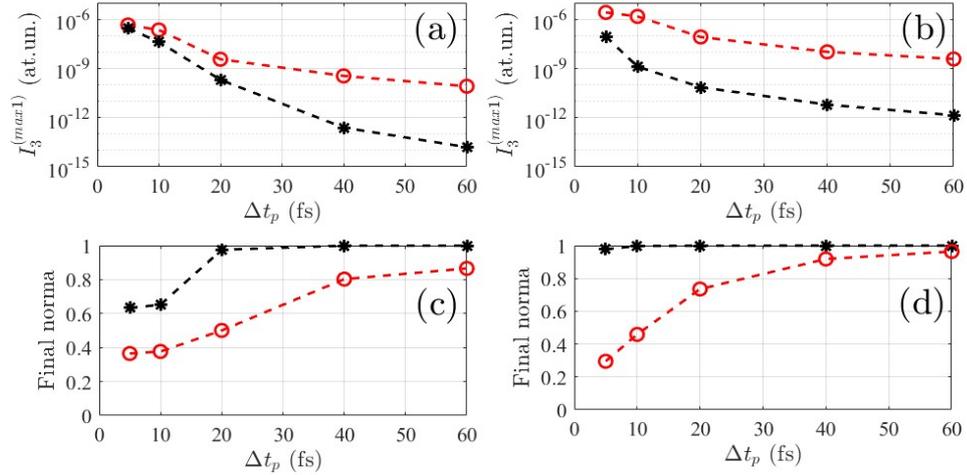

Fig. S7. Dependences (a,b) of the peak intensity of the third harmonic and (c,d) of the wave function norm within a sphere of radius $r_0 = 20$ a.u. at the end of the laser pulse on the laser pulse duration $\Delta t_p$ at the corresponding optimal value of its intensity $I_0$ (see markers in Fig. 3(a) in the main text of the article). Figures (a) and (c) refer to the Na atom, figures (b) and (d) to the $Mg^+$ ion. Black asterisks connected by dashed lines correspond to a nonresonant laser pulse, $\hbar\Omega = 2.5$ eV for Na and 7 eV for $Mg^+$. Red circles connected by dashed lines correspond to the resonant laser pulse, $\hbar\Omega = 1.99$ eV for Na and 4.77 eV for $Mg^+$. The data presented are obtained on the basis of a numerical solution of the TDSE from first principles with the corresponding effective one-electron potentials (see Section 1).

Let us illustrate the stability of the investigated regime of generation of ultrashort pulses of third harmonic to changes in the intensity of the laser field. Figure S8(a,b) shows the compression coefficient $\beta$ and the ratio of the peak intensities of the two most intense bursts in the time dependence of the third harmonic intensity, $I_3^{(max1)}/I_3^{(max2)}$, as a function of the laser pulse intensity $I_0$ at a fixed duration of $\Delta t_p = 20$ fs, obtained on the basis of the numerical solution of the TDSE. Figure S8(a) corresponds to the Na atom, and Fig. S8(b) to the $Mg^+$ ion. It is evident that the results of the first-principles TDSE solution are in qualitative agreement with the results of the analytical solution (2) and (3), shown for the Na atom and the $Mg^+$ ion in

Figs. S3 and S5, respectively. When the laser field intensity $I_0$ changes relative to its optimal (at a fixed laser pulse duration) value, first of all, the relative amplitude of the main and side bursts in the time dependence of the third harmonic changes, while the duration of the main burst changes insignificantly. Thus, this regime of generating ultrashort pulses of the third harmonic is stable to a change in the laser field intensity by 2-3 times.

Figure S8(c,d) shows the time dependences of the third harmonic intensity calculated based on the numerical solution of the TDSE at optimal parameters of the laser pulse. Figure S8(c) corresponds to the Na atom, and Fig. S8(d) to the $Mg^+$ ion. It is seen that the analytically predicted shape of the third harmonic pulse (see Figs. S4(c) and S6(c)) is in qualitative agreement with that obtained based on the TDSE numerical solution, and their differences, as was mentioned above when discussing Fig. 3 from the main text of the article, are explained by (a) the non-equidistance of the energies of the resonance states under consideration, (b) the involvement of an increasing number of bound states in the interaction with the field, and (c) ionization of the atom or ion.

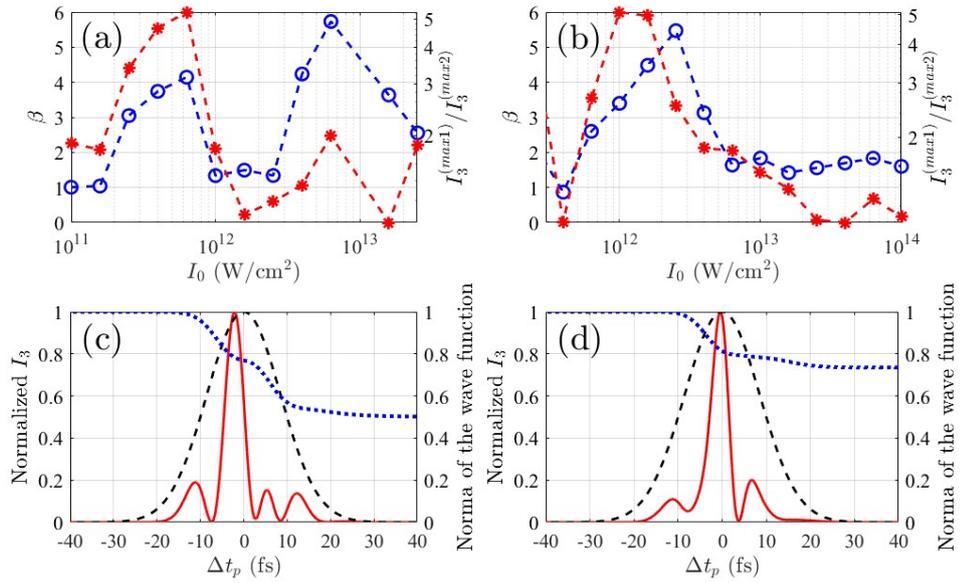

Fig. S8. (a,b) Dependences of the compression coefficient $\beta$ (left axis, blue circles connected by dashed lines) and the intensity ratio of the two most intense bursts (right axis, red stars with dashed lines) on the intensity of the laser pulse with a duration of $\Delta t_p = 20$ fs for (a) the Na atom and (b) the $Mg^+$ ion, calculated on the basis of the numerical solution of the TDSE from first principles (see Section 1). (c,d) Time dependences of the third harmonic intensity (left axis, red solid line) and the electron wave function norm (right axis, blue dotted line) for (c) the Na atom, $I_0 = 6.31 \times 10^{11}$ W/cm$^2$, $\Delta t_p = 20$ fs, $\hbar\Omega = 1.99$ eV, and (d) the $Mg^+$ ion, $I_0 = 1.59 \times 10^{12}$ W/cm$^2$, $\Delta t_p = 20$ fs, $\hbar\Omega = 4.77$ eV. The electron wave function norm was calculated within a sphere of radius $r_0 = 20$ a.u. for both the Na atom and the $Mg^+$ ion. Black dashed line is the laser field intensity envelope.

Finally, we note that the permissible frequency detuning of the laser field from the resonance with the two-photon transition, at which the effect of shortening of the third harmonic pulse is observed, is of the order of the effective Rabi frequency of the laser field, $DE_0$, and increases with decreasing its duration. Thus, for the $Mg^+$ ion with a laser pulse duration of 5 fs, the permissible range of carrier frequency tuning at the optimal intensity $I_0 = 2.5 \times 10^{13}$ W/cm$^2$ is ±10% in the vicinity of 4.77 eV (in the real $Mg^+$ ion, the transition energy is 4.43 eV), see the yellow lines in Fig. S9. In this case, the highest compression coefficient is reached at the optimal laser pulse intensity for frequency detunings of the laser field from the resonance of the order of the spectrum width of the laser pulse, see the dark green lines in Fig. S9.

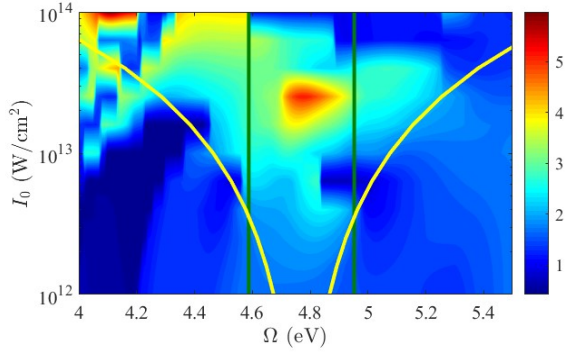

Fig. S9. The dependence of the compression coefficient $\beta$ of the third harmonic on the intensity $I_0$ and frequency $\Omega$ of the laser field, obtained on the basis of the numerical solution of the TDSE from first principles for the $Mg^+$ ion; the laser pulse duration is $\Delta t_p = 5$ fs. The dark green lines correspond to the spectral width of the Gaussian laser pulse with the duration $\Delta t_p = 5$ fs and the carrier frequency $\hbar\Omega = 4.77$ eV; the frequency interval between the yellow lines is equal to the effective Rabi frequency of the laser pulse, $DE_0$.

## 6. Parameters of resonance states of Na atom and $Mg^+$ ion

This section presents the characteristics of the resonance states of the Na atom (Table 1) and the $Mg^+$ ion (Tables 2, 3) calculated on the basis of the one-electron potential ("model atom", "model ion", "model") used to solve the TDSE from first principles (see Section 1). Also given are the corresponding characteristics of the real Na atom and $Mg^+$ ion taken from the NIST website [3].

It should be noted that, as follows from the presented data, the one-electron potentials used make it possible to reproduce with good accuracy the energies of the lower states of the Na atom and the $Mg^+$ ion and the dipole moments of the transitions between them.

Table 1. Parameters of resonance states of the Na atom

|  | Model | NIST |
|---|---|---|
| Energy of transition $\|1\rangle \leftrightarrow \|2\rangle$ ($\|2p^63s\rangle \leftrightarrow \|2p^63p\rangle$), $\hbar\omega_{21}$ (eV) | 1.99 | 2.10 |
| Energy of transition $\|2\rangle \leftrightarrow \|3\rangle$ ($\|2p^63p\rangle \leftrightarrow \|2p^65s\rangle$), $\hbar\omega_{32}$ (eV) | 1.94 | 2.01 |
| Energy of transition $\|2\rangle \leftrightarrow \|4\rangle$ ($\|2p^63p\rangle \leftrightarrow \|2p^64d\rangle$), $\hbar\omega_{42}$ (eV) | 2.11 | 2.18 |
| Ionization potential (eV) | 4.95 | 5.14 |
| Transition dipole moment modulus $\|1\rangle \leftrightarrow \|2\rangle$, $\|d_{12}\|$ (a. u.) | 2.5869 | 2.5014 |
| Transition dipole moment modulus $\|2\rangle \leftrightarrow \|3\rangle$, $\|d_{23}\|$ (a. u.) | 0.5319 | 0.4385 |
| Transition dipole moment modulus $\|2\rangle \leftrightarrow \|4\rangle$, $\|d_{24}\|$ (a. u.) | 0.8657 | 0.8558 |

Table 2. Parameters of resonance states of the Mg$^+$ ion

| | Model | NIST |
|---|---|---|
| Energy of transition $\|1\rangle \leftrightarrow \|2\rangle$ ($\|2p^63s\rangle \leftrightarrow \|2p^63p\rangle$), $\hbar\omega_{21}$ (eV) | 4.77 | 4.43 |
| Energy of transition $\|2\rangle \leftrightarrow \|3\rangle$ ($\|2p^63p\rangle \leftrightarrow \|2p^63d\rangle$), $\hbar\omega_{32}$ (eV) | 4.74 | 4.43 |
| Energy of transition $\|2\rangle \leftrightarrow \|4\rangle$ ($\|2p^63p\rangle \leftrightarrow \|2p^64s\rangle$), $\hbar\omega_{42}$ (eV) | 4.29 | 4.22 |
| Ionization potential (eV) | 15.63 | 15.04 |
| Transition dipole moment modulus $\|1\rangle \leftrightarrow \|2\rangle$, $\|d_{12}\|$ (a. u.) | 1.6067 | 1.6773 |
| Transition dipole moment modulus $\|2\rangle \leftrightarrow \|3\rangle$, $\|d_{23}\|$ (a. u.) | 1.7717 | 1.8588 |
| Transition dipole moment modulus $\|2\rangle \leftrightarrow \|4\rangle$, $\|d_{24}\|$ (a. u.) | 1.1580 | 0.9796 |

Table 3. Parameters of the states of the Mg$^+$ ion resonantly populated due to three-photon excitation

| | Model | NIST |
|---|---|---|
| Energy of transition $\lvert 2p^6 4s \rangle \leftrightarrow \lvert 2p^6 6p \rangle$ (eV) | 4.59 | 4.44 |
| Energy of transition $\lvert 2p^6 4s \rangle \leftrightarrow \lvert 2p^6 7p \rangle$ (eV) | 5.18 | 5.01 |
| Energy of transition $\lvert 2p^6 3d \rangle \leftrightarrow \lvert 2p^6 6p \rangle$ (eV) | 4.14 | 4.23 |
| Energy of transition $\lvert 2p^6 3d \rangle \leftrightarrow \lvert 2p^6 7p \rangle$ (eV) | 4.72 | 4.80 |
| Energy of transition $\lvert 2p^6 3d \rangle \leftrightarrow \lvert 2p^6 6f \rangle$ (eV) | 4.61 | 4.66 |
| Energy of transition $\lvert 2p^6 3d \rangle \leftrightarrow \lvert 2p^6 7f \rangle$ (eV) | 5.01 | 5.06 |
| Transition dipole moment modulus $\lvert 2p^6 4s \rangle \leftrightarrow \lvert 2p^6 6p \rangle$ (a. u.) | 0.0306 | 0.0183 |
| Transition dipole moment modulus $\lvert 2p^6 4s \rangle \leftrightarrow \lvert 2p^6 7p \rangle$ (a. u.) | 0.0337 | 0.0265 |
| Transition dipole moment modulus $\lvert 2p^6 3d \rangle \leftrightarrow \lvert 2p^6 6p \rangle$ (a. u.) | 0.0543 | 0.0883 |
| Transition dipole moment modulus $\lvert 2p^6 3d \rangle \leftrightarrow \lvert 2p^6 7p \rangle$ (a. u.) | 0.0341 | 0.0556 |
| Transition dipole moment modulus $\lvert 2p^6 3d \rangle \leftrightarrow \lvert 2p^6 6f \rangle$ (a. u.) | 0.4720 | 0.4456 |
| Transition dipole moment modulus $\lvert 2p^6 3d \rangle \leftrightarrow \lvert 2p^6 7f \rangle$ (a. u.) | 0.3150 | 0.3068 |